\documentclass[10pt,
               aps,
               pre,
               twocolumn,
               showpacs, 
               groupedaddress]{revtex4-2}
\usepackage[utf8]{inputenc}
\usepackage{graphicx}
\usepackage{float}
\usepackage{caption}
\usepackage{ragged2e}
\usepackage[dvipsnames,table]{xcolor}
\usepackage[caption=false]{subfig}
\usepackage{amsmath, amsfonts, amssymb}
\usepackage{natbib}
\usepackage[linktocpage=true,
  colorlinks=true, 
  pdfborder={0 0 0},
  linkcolor=blue,
  citecolor=blue,
  filecolor=yellow,
  urlcolor=blue,
  bookmarks,
  pdfauthor={},
]{hyperref}
\usepackage{tikz}
\graphicspath{{images/}}

\makeatletter
\newcommand*{\rom}[1]{\expandafter\romannumeral #1}
\makeatother

\begin{document}

\title{Active Ornstein-Uhlenbeck particle under stochastic resetting}
\author{Uma Shankari}
\affiliation{Department of Physics, University of Kerala, Kariavattom, Thiruvananthapuram-$695581$, India}


\author{Mamata Sahoo}
\email{jolly.iopb@gmail.com}
\affiliation{Department of Physics, University of Kerala, Kariavattom, Thiruvananthapuram-$695581$, India}

\date{\today}
\begin{abstract}
We investigate the dynamics of an inertial active Ornstein-Uhlenbeck (OU) particle in the presence of stochastic resetting. 
Using renewal approach, we compute the mean square displacement (MSD) and position probability distribution functions both in the overdamped and underdamped regimes, while suspended in a viscous and viscoelastic environment. In contrast to the normal active OU particle, the MSD of an active OU particle with resetting displays an initial diffusive and long-time or steady-state non-diffusive behavior. The steady-state MSD has a dependence both on the resetting rate and on the persistent duration of activity. It gets suppressed with an increase in the resetting rate and approaches zero for an infinitely large resetting rate. Moreover, steady-state MSD has a nonmonotonic impact on the duration of activity. For a very short and very long persistent duration of activity, the system behaves like a passive Brownian system. However, for an intermediate range of activity time, the steady-state MSD is enhanced, allowing the particle to explore a larger region, thereby increasing the probability of encountering a wide-spread target. These results are further supported by the position probability distribution function, which initially grows wide, shows a maximum spread, and narrows with further increase in the duration of activity. Moreover, when the particle is suspended in a viscoelastic medium characterized by the presence of finite memory, the MSD interestingly develops an intermediate-time plateau and the plateau depends on the strength of the viscoelasticity and the viscoelastic relaxation time. With an increase in strength of the viscoelasticity, the width of the plateau increases and the MSD suppresses. In addition,  slow viscoelastic relaxation makes memory effects weaker and hence the intermediate plateau disappears and steady state MSD gets enhanced, implying low strength and longer persistence of memory are beneficial for a wide spread target search. Similarly, for slow resetting, the intermediate plateau is more distinct, whereas fast resetting results in the system approaching steady state faster, overcoming the intermediate plateau in MSD. This emerging behavior might be due to the complex interplay between the resetting mechanism, activity, and memory effects. Finally, our analytical results are in good agreement with numerical simulation.

\end{abstract}

\maketitle
\section{INTRODUCTION}\label{sec:intro}
Stochastic resetting is the process in which the evolution of a system is interrupted and reset to its initial state at random intervals. It was first introduced in the diffusion of a Brownian particle \cite{evans2011diffusion}, with the aim of optimizing search processes. Search processes involving diffusive behavior are common in natural systems such as foraging animals \cite{benichou2005optimal,bartumeus2009optimal,viswanathan2011physics}, intracellular transport \cite{bressloff2020modeling}, and protein molecules binding to DNA \cite{berg1981diffusion,coppey2004kinetics}. In passive Brownian systems, stochastic resetting drives the system out of equilibrium by altering its temporal relaxation dynamics, leading to a non-equilibrium steady state \cite{evans2011diffusion,biswas2025resetting}. In general, the mean first passage time (MFPT) of a freely diffusing Brownian particle is infinite. However, the introduction of stochastic resetting provides a finite MFPT, and therefore there exists an optimal resetting rate for the minimal MFPT, making it an efficient tool for search optimization \cite{evans2011diffusion_with_opt_resetting,gupta2022stochastic,evans2020stochastic}.  


Active matter systems provide a fundamental framework for modelling and understanding of non-equilibrium physics and autonomous motion in biology, with wide applications in material science, medicine and robotics. 
In contrast to the passive Brownian systems which deal with random thermal motion caused by equilibrium fluctuations, 
active matter is composed of self-propelled particles that exhibit autonomous motion by consuming energy from the environment, thereby driving the system out of equilibrium \cite{de2015introduction,bechinger2016active,gompper20202020,dombrowski2004self} and making them applicable to more realistic situations. For example, chemically driven motion of microorganisms \cite{berg1972chemotaxis,jones2021stochastic}, intracellular transport \cite{bressloff2020modeling} in living bodies, drug-delivery \cite{park2025biohybrid} and movement of micro-robots engineered in laboratories \cite{nauber2023medical,liu2023responsive}. 


Considering the established effects of stochastic resetting in the passive Brownian systems, it is natural and intriguing to analyze the impact of resetting in self-propelled active systems which are inherently more complex and closer to the real-world applications. Active matter with resetting has been explored in different contexts. 
In one dimension, resetting the position and reversing the direction of self-propelled velocity of an active transport or for a run and tumble particle is observed to reduce the overall search time \cite{evans2018run,bressloff2020modeling}. In addition, when a run and tumble particle is suspended in a thermal environment, it outperforms the passive Brownian search for low noise strength and high self-propelled velocity \cite{pal2024active}. 
In two dimensions, resetting the position and orientation of an active Brownian particle leads to a stationary state \cite{kumar2020active,shee2025steering}. On the other hand, in the case of only orientation reset, the system remains diffusive \cite{kumar2020active,baouche2024active} and maintains the anisotropic properties even in long time regimes, for an anisotropic active Brownian particle \cite{ghosh2025anisotropic}. Moreover, resetting the position and orientation of such an active particle while confined by a harmonic trap displays a transition from the resetting dominated regime to the activity dominated regime with an increase in the trap stiffness \cite{shee2025active}. Similarly, a charged self-propelled particle with stochastic resetting and in the presence of an external magnetic field displays a non-trivial stationary state and prolongs the time taken to reach the target when compared to its passive counterpart \cite{abdoli2021stochastic}. 
Alike a passive Brownian particle, an active Brownian particle with resetting also leads to a finite MFPT with an optimal resetting rate \cite{scacchi2018mean}. However, the minimum MFPT depends on the initial orientation of the particle \cite{baouche2025optimal}.

Although the effects of stochastic reset have been studied for run and tumble and active Brownian particle, the effects of reset on the dynamics of an active OU particle \cite{lehle2018analyzing,bonilla2019active,martin2021statistical,caprini2021inertial}, 
is not widely explored. In this work, we are mainly interested in examining the dynamical behavior of an active OU particle with stochastic resetting using both the analytical approach and numerical simulation. Our main focus is to investigate the impact of complex interplay of activity and resetting on the dynamics, while subjected to both a viscous and a viscoelastic suspension. We observe that the steady-state MSD shows a nonmonotonic dependence on the duration of activity. For very short and long activity timescales, the impact of activity on the steady-state MSD is negligible, whereas for intermediate ranges of the activity timescales, the steady-state MSD is enhanced, implying the encounter of a widespread target. While suspended in a viscoelastic environment characterized by the presence of finite memory, the MSD develops an intermediate-time plateau and the plateau regime depends both on the strength of the viscoelasticity and the relaxation time scale. With an increase in viscoelastic strength, the plateau becomes wider and steady-state MSD is suppressed. However, increasing the relaxation time delays the onset of the plateau and reduces its width. The plateau regime is prominent for the slower resetting mechanism and gradually decreases with increasing resetting rate.


\section{MODEL AND METHOD}\label{sec:model}

We consider the motion of a free active OU particle of mass $m$ in a viscous fluid with viscous coefficient $\gamma_f$, driven by 
a Gaussian white noise $\eta(t)$. The Langevin's equation of motion of the particle \cite{caprini2021inertial,muhsin2022inertial,arsha2023steady,arsha2024inertial} is given by
\begin{equation}\label{eq:lang_eqn_wo_r}
    m\dot v = -\gamma_f \dot x + f_a(t) + \eta(t).
\end{equation}
Here, $x$ and $v$ denote the position and velocity coordinate of the particle. The left-hand side of Eq. \eqref{eq:lang_eqn_wo_r} represents the inertial force and the first term on the right-hand side represents the viscous force. $\eta(t)$ is a delta-correlated Gaussian white noise with the following statistical properties.
\begin{equation}\label{white_noise_stat_prop}
    \langle \eta(t) \rangle =0 \quad; \quad \langle \eta(t) \eta(s) \rangle=2\gamma_f k_BT\delta(t-s),
\end{equation}
with $k_B$ as the Boltzmann constant and $T$ is the temperature of the medium. The angular bracket $\langle\cdots\rangle$ denotes the ensemble average over noise. $f_a(t)$ is the active force with strength $f_0$ that follows the Ornstein-Uhlenbeck process as \cite{nguyen2021active,muhsin2023inertial,muhsin2025active}
\begin{equation}\label{eq:aoup_noise_evolution}
    t_c \dot f_a(t) = -f_a(t) + f_0 \sqrt{2 t_c} \  \zeta(t). 
\end{equation}
The statistical properties of $f_a(t)$ are
\begin{equation}\label{aoup_noise_stat_prop}
    \langle f_a(t)\rangle=0\quad;\quad\langle f_a(t) f_a(s)\rangle=f_0^2\exp{\left(-\frac{|t-s|}{t_c}\right)}.
\end{equation}
Here, the correlation decays exponentially with a time constant $t_c$ and hence $t_c$ is the time upto which there exists a finite correlation in the medium.  It quantifies the duration of activity or self-propulsion in the medium. Hence, $t_c$ can be referred to as the activity timescale. $\zeta(t)$ is a Gaussian white noise with properties $\langle \zeta \rangle=0 \ ; \ \langle\zeta(t) \zeta(s)\rangle=\delta(t-s)$.   

The dynamics of the particle can be described by the following stochastic equations
\begin{equation*}
    \dot x=v,
\end{equation*}
\begin{equation*}
    \dot v=-\frac{\gamma_f}{m}\dot x+\frac{1}{m}f_a(t)+\frac{1}{m}\eta(t), 
\end{equation*}
and
\begin{equation}\label{eq:ud_dynamics}
    \dot f_a= -\frac{1}{t_c}f_a(t)+f_0 \sqrt{\frac{2}{t_c}}\zeta(t).
\end{equation}

Now, by introducing the resetting mechanism to the system, the entire dynamics of the particle is reset to its initial state at stochastic intervals. The reset rate is denoted by the parameter $r$. At any instant of time $t$, the particle has the following options for its evolution during the subsequent infinitesimal time interval $dt$. 
\begin{equation}\label{r_dynamics}
    \textbf{X}(t+dt)= \begin{cases}
        \textbf{X}(0) \quad \quad \quad \quad \quad \ \ $with probability$ \ rdt \\
        Eq.\hspace{0.1cm}\eqref{eq:ud_dynamics} \ \quad \quad \quad $with probability$\ 1 - rdt. \\ 
    \end{cases}    
\end{equation}
Here, $\textbf{X}(t)=(x(t),v(t),f_a(t))^T$ and $\textbf{X}(0)$ is the initial configuration of the system. \textit{i.e.,}$\textbf{X}(0)=(x_0,v_0,f_{a0})^T$. $r$ is a dynamic parameter with a constant value in the range $0\leq r \leq \infty $. From Eq. \eqref{r_dynamics}, it is evident that the resetting dynamics preserves the Markovian nature of the system. 

In this work, we primarily investigate the transport properties of an active OU particle with stochastic resetting in both viscous and viscoelastic environment. To gain insights into its dynamics, we compute the mean square displacement (MSD), probability distribution functions with the help of renewal approach and examine the behaviour across different time regimes of the dynamics. 
The MSD of the particle can be calculated as
\begin{equation}\label{eq.msd0_formula}
    \langle \Delta x^2(t) \rangle=\langle(x(t)-x(0))^2\rangle.
\end{equation}

\subsubsection*{The Renewal Approach}
The reset mechanism introduced here is a renewal process, in which the dynamics of the system is renewed at a Poisson rate $r$~\cite{evans2011diffusion}. Therefore, the position probability distribution $P_r(x,t)$ of the particle with resetting can be written in terms of the position probability distribution without resetting $P_0(x,t)$ using the following renewal equation~\cite{evans2020stochastic,trajanovski2023ornstein}.
\begin{equation}\label{eq.renewal_pdf}
    P_r(x,t)= e^{-rt} P_0(x,t) + \int_0^t r e^{-rt'} P_0(x,t')dt'.
\end{equation}
In Eq. \eqref{eq.renewal_pdf}, the first term of right hand side corresponds to the probability of events that do not reset for the entire time interval $t$. The second term corresponds to the probability of events in which resetting occurs for a small time interval $dt'$ after a time $t'$~\cite{evans2020stochastic}. 
The moments of the distribution can also be calculated using a renewal equation similar to Eq. \eqref{eq.renewal_pdf}. The MSD with resetting $\langle  \Delta x^2(t)\rangle_r$ can be obtained from the MSD without resetting $\langle \Delta x^2(t)\rangle_0$ in the following way \cite{kumar2020active,trajanovski2023ornstein}
\begin{equation}\label{eq.msd_renewal_form}
   \langle\Delta x^2(t)\rangle_r = e^{-rt} \langle \Delta x^2(t)\rangle_0 + \int_0^t r e^{-rt'} \langle\Delta x^2(t')\rangle_0 dt'.
\end{equation}

Using Euler-Maruyama method \cite{platen1992numerical}, the dynamics described in Eq. \eqref{r_dynamics} is simulated upto $10^4$ time steps. The numerical integrations in Eq. \eqref{eq.renewal_pdf} and Eq. \eqref{eq.msd_renewal_form} is performed with time steps $dt=10^{-5}$. All the physical quantities are averaged over $10^4$ independent realisations.
 
\section{RESULTS AND DISCUSSION}\label{sec:result}
\subsection{In a viscous medium}
\subsection*{Without resetting}
It is to be noted that when the viscous coefficient $\gamma_f$ in Eq. \eqref{eq:lang_eqn_wo_r} is sufficiently large, the viscous force outweighs the inertial effects 
and the particle follows the dynamics 
\begin{equation}\label{od_xdot}
    \dot x=\frac{1}{\gamma_f}f_a(t)+\frac{1}{\gamma_f}\eta(t).
\end{equation}

In the absence of resetting, the probability distribution of the particle at any time $t$, $P_0(x, f_a, t)$ follows the Fokker-Planck equation (FPE) \cite{risken1996fokker,balakrishnan2008elements}
\begin{equation}\label{od_fpe_wo_r}
    \begin{split}
        \frac{\partial P_0}{\partial t}=\frac{\partial}{\partial x} \left(-\frac{f_a}{\gamma_f}+\frac{k_BT}{\gamma_f}\frac{\partial}{\partial x}\right)P_0\\
        +\frac{\partial}{\partial f_a}\left(\frac{f_a}{t_c}+\frac{f_0^2}{t_c}\frac{\partial}{\partial f_a}\right)P_0
    \end{split}  
\end{equation}
with initial condition $P_0(x,f_a,0)=\delta(x-x_0)\delta(f_a-f_{a0})$.
    
The solution to the linear FPE in Eq. \eqref{od_fpe_wo_r} has the following form \cite{van1992stochastic} 
\begin{equation}\label{eq.od_v_p0_exp}
    P_0(x,t)= \frac{1}{\sqrt{2 \pi \sigma}}\exp\left\{{-\frac{\Big[x(t)-\langle x(t)\rangle\Big]^2}{2\sigma}}\right\}.    
\end{equation}
Here, $\sigma=\langle\Delta x^2(t)\rangle_0 - \langle\Delta x(t)\rangle^2_0$. Integrating, Eq. \eqref{od_xdot} and taking the initial position $x_0=0$, we obtain the mean displacement to be zero, $i.e., \langle\Delta x(t) \rangle_0=0$. Using Eq. \eqref{eq.msd0_formula}, with the help of the correlation matrix method~\cite{van1992stochastic}, as explained in appendix \ref{app.correlation_matrix}, 
the MSD can be calculated as 

\begin{equation}\label{od_v_msd_full_exp}
    \langle\Delta x^2(t) \rangle _0=\frac{2k_BT t}{\gamma_f}+\frac{f_0^2 t_c\left\{2 t+\left[(4 e^{-\frac{t}{t_c}}-e^{-\frac{2t}{t_c}})-3\right]t_c\right\}}{\gamma_f^2}.
\end{equation}

In the limit $t\to 0$, $\langle\Delta x^2(t) \rangle_0$ follows $\frac{2k_BT}{\gamma_f}t$, while in the limit $t\to \infty$, it approaches 
\begin{equation}\label{eq.od_v_msd0_tlong}
    \underset{ t \to \infty}{\lim}\langle\Delta x^2(t)\rangle_0=\frac{2(\gamma_f k_BT+f_0^2 t_c)}{\gamma_f^2} t.
\end{equation}
Equation \eqref{eq.od_v_msd0_tlong} indicates that the influence of activity is significant only at long times. In the long time limit, the $\langle\Delta x^2(t) \rangle_0$ remains proportional to $t$, implying that the particle continues to diffuse indefinitely. 

With the help of correlation matrix method as explained in appendix \ref{app.correlation_matrix}, $\sigma$ can be calculated and substituting $\sigma$ in Eq. \eqref{eq.od_v_p0_exp}, we get the exact expression of $P_0(x,t)$. 

\begin{figure}[h]
    \centering
    \includegraphics[width=\linewidth]{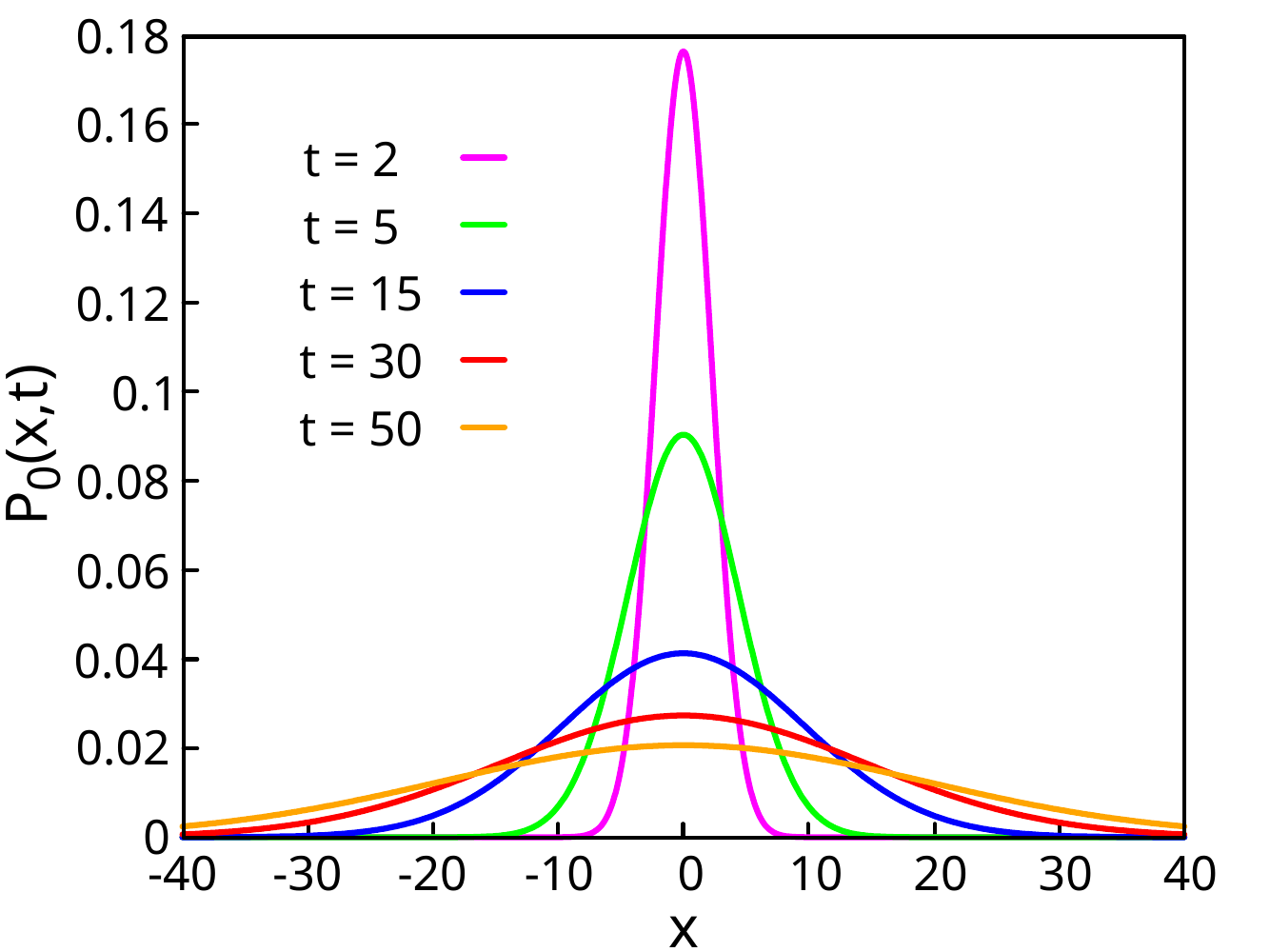}
    \caption{\justifying$P_0(x,t)$ of an overdamped particle as a function of $x$ for different values of $t$. The common parameters are $\gamma_f=f_0=k_BT=1.0$ and $t_c=3.0$}
    \label{fig:od_v_p0_t}
\end{figure}

In Fig. \ref{fig:od_v_p0_t}, we have plotted $P_0(x,t)$ for different time instances. At any time $t$, the distribution retains a Gaussian form and gradually spreads as time progresses, indicating that the probability of finding the particle farther from its initial position increases with time. 
This aligns with the long-time limit behavior of MSD [Eq. \eqref{eq.od_v_msd0_tlong}]. 

In the underdamped version of the dynamics, the influence of inertia is significant and the dynamics is described by Eq. \eqref{eq:ud_dynamics}. The following FPE describes the evolution of the corresponding probability distribution $P_0(x,v,f_a,t)$ for an inertial active OU particle
\begin{widetext}
\begin{equation}\label{ud_fpe_wo_r}
    \begin{split}
        \frac{\partial P_0}{\partial t}=-v\frac{\partial P_0}{\partial x}+\frac{\partial}{\partial v}\left(\frac{v}{t_m}-\frac{f_a}{m}+\frac{k_BT}{\gamma_f t_m^2}\frac{\partial}{\partial v}\right)P_0+\frac{\partial}{\partial f_a}\left(\frac{f_a}{t_c}+\frac{f_0^2}{t_c}\frac{\partial}{\partial f_a}\right)P_0.
    \end{split}
\end{equation}   
\end{widetext}
Here, $t_m=\frac{m}{\gamma_f}$ represents the inertial timescale. When $\gamma_f$ is very large, $t_m \to 0$ and the system approaches the overdamped limit. We have exactly solved the Eq. \eqref{ud_fpe_wo_r} using the correlation matrix method as described in appendix \ref{app.correlation_matrix} and the solution $P_0(x,v,f_a,t)$ takes a Gaussian form given in Eq. \eqref{ud_v_prob_x_xi_t}.  
Now, integrating $P_0(x,v,f_a,t)$ over other variables, we get the $P_0(x,t)$ as the same form of Eq. \eqref{eq.od_v_p0_exp}. 

However, the mean displacement can be calculated as $\langle\Delta x(t) \rangle_0=v_0 t_m\big(1-e^{-t/t_m}\big)$~\cite{nguyen2021active}. Similarly, the MSD without resetting can be calculated as

\begin{widetext}

    \begin{equation}\label{eq.ud_v_msd0}
    \begin{split}
        \langle\Delta x^2(t) \rangle_0 &= v_0^2t_m^2\Big(1-e^{-\frac{t}{t_m}}\Big)^2+\frac{k_BT}{\gamma_f}\Big[2t+t_m\big(4e^{-\frac{t}{t_m}}-e^{-\frac{2t}{t_m}}-3\big)\Big]
        +\frac{f_0^2t_c}{\gamma^2\alpha_-^2\alpha_+}\Bigg[\alpha_-^2\Big(\beta+2t\alpha_+-3\alpha_+^2\Big)\\
        &\hspace{0.5 cm}-\alpha_+\Big(t_c^3\ e^{\frac{-2t}{t_c}}+t_m^3\ e^{\frac{-2t}{t_m}}\Big)-4\alpha_+\alpha_-\Big(t_m^2\ e^{\frac{-t}{t_m}}-t_c^2\ e^{\frac{-t}{t_c}}-\beta^2e^{-\frac{t\alpha_+}{\beta}}\Big)\Bigg],
    \end{split}   
\end{equation}
\end{widetext}
where $\alpha_+=t_c+t_m$, $\alpha_-=t_c-t_m$ and $\beta=t_c t_m$. In the limit $t \to 0$, $\langle\Delta x^2(t) \rangle_0$ [Eq. \eqref{eq.ud_v_msd0}] tends to $v_0^2t^2+\left( \frac{2k_BT}{3 t_m^2\gamma_f}-\frac{v_0^2}{tm} \right)t^3$, reflecting the initial ballistic motion of the particle. 
In the long-time limit, that is, for the $t \to \infty$ limit, $\langle\Delta x^2(t)\rangle_0$ becomes
\begin{equation}\label{eq.ud_v_msd0_st}
\begin{split}
    \underset{t \to \infty}{\lim} \langle\Delta x^2(t) \rangle_0&=2\Bigg[\frac{ k_BT}{\gamma_f}+\frac{ f_0^2 t_c}{\gamma_f^2}\Bigg]t.
\end{split}
\end{equation}

From the above equation, it is to be noted that $\langle x^2(t)_0$ is proportional to $t$, which confirms the long time diffusive nature of the particle.

\subsection*{With resetting}
\quad When an overdamped active OU particle is subjected to stochastic resetting with resetting rate $r$, the particle evolves according to Eq. \eqref{od_xdot}, having the possibility of resetting its dynamics to the initial state. Now, the time evolution of the probability distribution with resetting, $P_r(x, f_a, t)$ is given by~\cite{evans2011diffusion,gupta2022stochastic,trajanovski2023ornstein}
\begin{widetext}
\begin{equation}\label{od_fpe_r}
    \begin{split}
        \frac{\partial P_r}{\partial t}&=\frac{\partial}{\partial x} \left(-\frac{f_a}{\gamma_f}+\frac{k_BT}{\gamma_f}\frac{\partial}{\partial x}\right)P_r+\frac{\partial}{\partial f_a}\left(\frac{f_a}{t_c}+\frac{f_0^2}{t_c}\frac{\partial}{\partial f_a}\right)P_r-rP_r+r\delta(x-x_0).
    \end{split}  
\end{equation}
The MSD with reset $\langle\Delta x^2(t)\rangle_r$ can be obtained by substituting $\langle \Delta x(t)^2\rangle_0$ in the renewal equation [Eq. \eqref{eq.msd_renewal_form}] and it takes the form 
    \begin{equation}\label{eq.od_msdr_full_exp}
    \begin{split}
        \langle\Delta x^2(t)\rangle_r=\frac{2k_BT(1-e^{-rt})}{r\gamma_f}+\frac{2 f_0^2 t_c}{r \gamma_f^2}\Bigg\{\frac{2\big(1-e^{-rt}\big)}{\big(1+rt_c\big)\big(2+rt_c\big)}+rt_c\Bigg(e^{-\frac{t}{t_c}}\Bigg[\frac{e^{(1+rt_c)}}{1+rt_c}-\frac{e^{(2+rt_c)}}{2+rt_c}\Bigg]-e^{-rt}\Big[3+rt_c\Big]\Bigg)\Bigg\}.  
    \end{split}
\end{equation}
\end{widetext}

In the limit $r \to 0$, the $\langle\Delta x^2(t) \rangle_r$ becomes $\langle\Delta x^2(t) \rangle_0$ [Eq. \eqref{od_v_msd_full_exp}]. Similarly, for large resetting rate, the particle is constantly pulled back to the initial position, making it difficult for the particle to move very far away from the initial position. Therefore, in the limit $ r \to \infty$, $\langle\Delta x^2(t) \rangle_r$ approaches zero. In $t \to 0$ limit, $\langle\Delta x^2 (t) \rangle_r$ grows as $\frac{2k_BT}{\gamma_f} t$ and hence has no impact of resetting or activity. 
The steady-state MSD, $\langle\Delta x^2(t) \rangle_r^{st}$ is found to be
\begin{equation}\label{eq.od_v_msdr_st}
\begin{split}
    \langle\Delta x^2(t) \rangle_r^{st}&=\underset{t \to \infty}{\lim }\langle\Delta x^2(t)\rangle\\
    &=\frac{2k_BT}{r\gamma_f}+\frac{2 \left(2 \hspace{0.05cm} f_0^2 \hspace{0.05cm} t_c \right)}{r \gamma_f^2 \left(1+rt_c \right)\left( 2+rt_c \right)}.
\end{split}
\end{equation}
From the above equation, it is confirmed that 
the first term is due to the impact of thermal effects and the second term is due to the influence of activity on the steady state MSD. In $t_c \to 0$ limit, the second term vanishes and the MSD is same as the case of a passive Brownian  particle. The lower value $t_c$ results in the persistence of activity for a short interval of time, and hence the impact of activity on MSD is less significant. However, in the limit $t_c \to \infty$, the second term of Eq. \eqref{eq.od_v_msdr_st} vanishes. This is because for a very large $t_c$ value, the strength of the active force is weakened, and as a result the impact of activity on MSD decreases. Therefore, for very low and very high values of $t_c$ the impact of activity is not significant and for intermediate range of $t_c$, the activity enhances steady-state MSD.

\begin{figure}[h]
    \centering
    \includegraphics[width=\linewidth]{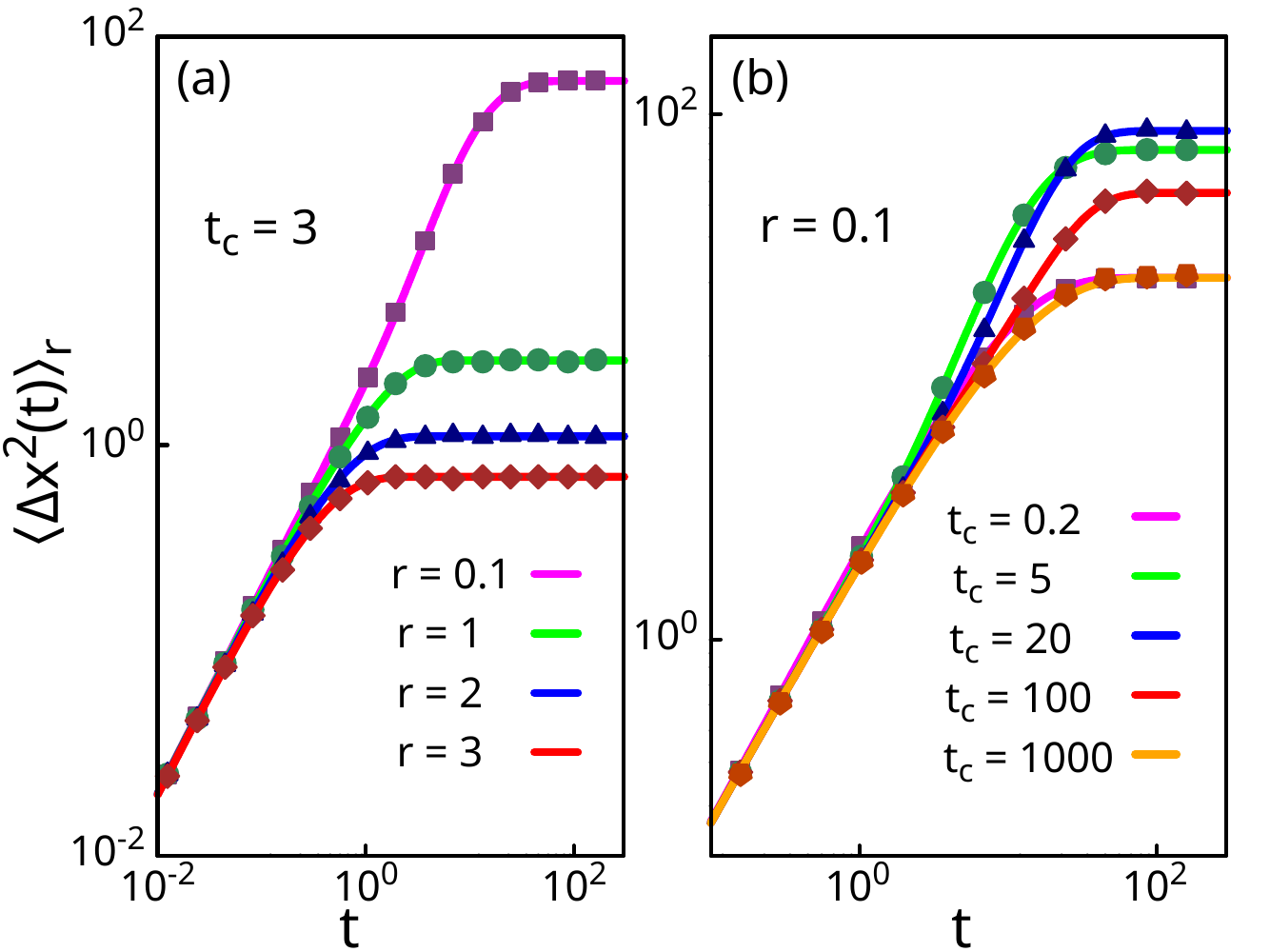}
    \caption{\justifying$\langle\Delta x^2(t) \rangle_r$ of an overdamped particle as a function of $t$ for different values of $r$ in $(a)$ and for different values of $t_c$ in $(b)$. The other common parameters are $\gamma_f=f_0=k_BT=1.0$. The solid lines represent analytical results and the markers represent the results from simulation.}
    \label{fig:od_v_msd_r_tc}
\end{figure}

In Fig. \ref{fig:od_v_msd_r_tc}, we have plotted $\langle \Delta x^2(t) \rangle_r$ as a function of $t$, for different values of $r$ in Fig. \ref{fig:od_v_msd_r_tc}(a) and for different values of $t_c$ in Fig. \ref{fig:od_v_msd_r_tc}(b), respectively. For a fixed value of $r$ and $t_c$, initially the MSD grows linearly with time and saturates to a constant steady-state value in the long time limit. With an increase in the value of $r$, the MSD attains the steady state value faster. At the same time, the steady-state MSD is suppressed with increase in $r$ value [see Fig. \ref{fig:od_v_msd_r_tc}(a)]. Similarly, with an increase in the value of $t_{c}$, it is observed that the initial time regimes of MSD are not affected [Fig. \ref{fig:od_v_msd_r_tc}(b)]. However, steady-state MSD initially increases with $t_{c}$, approaches a maximum value, and then decreases and saturates. Therefore, steady-state MSD has a nonmonotonic impact on the duration of activity.

Substituting $P_0(x,t)$ for an overdamped particle from Eq. \eqref{eq.od_v_p0_exp} in the renewal equation [Eq. \eqref{eq.renewal_pdf}], we have numerically evaluated the distribution function with resetting $P_r(x,t)$. 
\begin{figure*}[t]
    \centering
    \includegraphics[width=\linewidth]{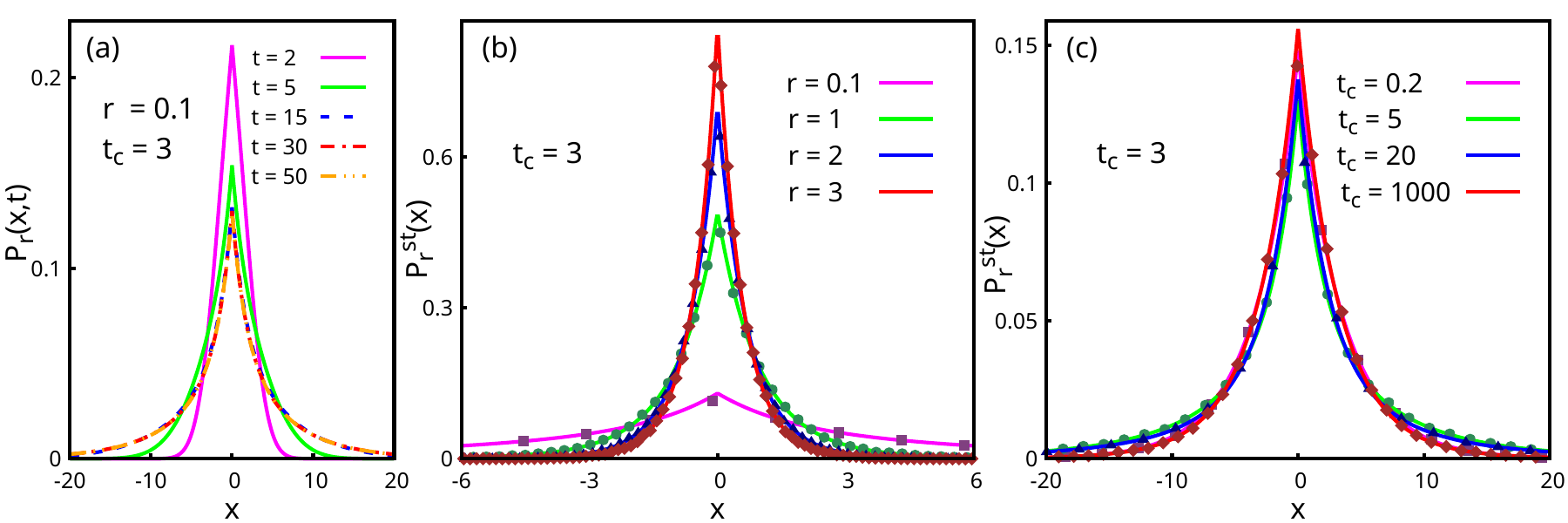}
    \caption{\justifying$P_r(x,t)$ of an overdamped particle plotted as a function of $x$ for different values of $t$ in $(a)$, $P_r^{st}(x)$ as a function of $x$ for different values of $r$ in $(b)$, and for different values of $t_c$ in $(c)$, respectively. The other common parameters are $\gamma_f=f_0=k_BT=1.0$. The solid lines represent analytical results and the markers represent the results from numerical simulation.}
    \label{fig:od_v_pr_t_r_tc}
\end{figure*}
In Fig. \ref{fig:od_v_pr_t_r_tc}, we have plotted this $P_{r}(x,t)$ along with the simulated curves as a function of $x$ for different time instances, the steady-state position probability distribution $P_r^{st}(x)$ as a function of $x$ for different values of $r$ and $t_c$ in Figs. [\ref{fig:od_v_pr_t_r_tc}(a)$-$\ref{fig:od_v_pr_t_r_tc}(c)], respectively. From Fig. \ref{fig:od_v_pr_t_r_tc}(a), it is confirmed that at any instant of time, $P_r(x,t)$ takes the form of a non-Gaussian distribution function, especially exponential in nature with centre at the resetting point ($x_0=0$). 
The exponential nature of $P_r(x,t)$ implies that the probability of finding the particle near the reset point is significant and decreases exponentially as we move away from the resetting point. The resetting mechanism indicates that even at longer times, there is a non-zero probability current flowing towards the resetting point. This violates the principle of detailed balance condition, and as a result of this, the system reaches a non-equilibrium steady state \cite{gupta2022stochastic}. Therefore, for a longer instant of time, the distribution does not show any change in behavior [Fig. \ref{fig:od_v_pr_t_r_tc}(a)]. 
From Figs. [\ref{fig:od_v_pr_t_r_tc}(b) and \ref{fig:od_v_pr_t_r_tc}(c)], it is observed that for a fixed value of $r$ or $t_c$, $P_r^{st}(x)$ follows the exponential distribution centered at the origin. Furthermore, with increasing $r$ value, the distribution becomes narrower and, at the same time, the maximum value at the center increases [Fig. \ref{fig:od_v_pr_t_r_tc}(b)]. This implies that frequent resetting enhances the probability at the resetting point. However, for a fixed $r$, initially with an increase in the value of $t_{c}$, the distribution becomes wider and the peak suppresses. A further increase in $t_{c}$ results in a narrow distribution and a reduction in the peak value [see Fig. \ref{fig:od_v_pr_t_r_tc}(c)]. This complements the nonmonotonic impact of steady-state MSD on $t_c$.

\quad Similarly, in the presence of resetting, an underdamped active OU particle evolves according to the Eq. \eqref{eq:lang_eqn_wo_r} and intermittently resets to its initial state with a resetting rate $r$. The evolution of the corresponding probability distribution $P_r(x,v,f_a,t)$ is described by the following FPE 
\begin{widetext}
    \begin{equation}\label{ud_fpe_r}
    \begin{split}
        \frac{\partial P_r}{\partial t}=-v\frac{\partial P_r}{\partial x}+\frac{\partial}{\partial v}\left(\frac{\gamma_f v}{m}-\frac{f_a}{m}+\frac{\gamma_f}{m^2}\frac{\partial}{\partial v}\right)P_r
        +\frac{\partial}{\partial f_a}\left(\frac{f_a}{t_c}+\frac{f_0^2}{t_c}\frac{\partial}{\partial f_a}\right)P_r-rP_r+r\delta(x-x_0).
    \end{split}
\end{equation}

Substituting $\langle\Delta x^2(t) \rangle_0$ from Eq. \eqref{eq.ud_v_msd0} in Eq. \eqref{eq.msd_renewal_form}, we evaluate the MSD of an underdamped active OU particle with resetting as

    \begin{equation}\label{eq.ud_msdr_full_exp}
        \begin{split}
           \langle\Delta x^2 (t) \rangle _r &= \Bigg[\frac{2\big(2 k_BT + r t_m^2v_0^2\gamma_f\big)}{r\gamma_f \big(1+rt_m\big)\big(2+rt_m\big)} \Bigg]+e^{-rt}\Bigg\{2t_m\Bigg[\frac{\big(2 k_BT - t_mv_0^2\gamma
           _f\big)\ e^{-\frac{t}{t_m}}}{\gamma_f(1+rt_m)}-\frac{\big(k_BT-t_mv_0^2\gamma_f \big)\ e^{-\frac{2t}{t_m}}}{\gamma_f\big(2+rt_m \big)} \Bigg]-\frac{2 k_BT}{r\gamma_f} \Bigg\}\\
           &\hspace{0.5 cm}+\frac{1}{\gamma_f}f_0^2t_c\Bigg\{\frac{2\big(1- e^{-rt}\big)}{r}+\frac{\beta- 3 \alpha_+^2}{\alpha_+}-\frac{e^{-rt}}{\alpha_-^2 \alpha_+ }\Bigg[4\alpha_+\alpha_-\Big(t_m^2\ e^{-\frac{t}{t_m}}-t_c^2\ e^{-\frac{t}{t_c}}\Big) +\alpha_+\Bigg[t_c^3\ e^{-\frac{2t}{t_c}}+t_m^3\ e^{-\frac{2t}{t_m}}\Bigg]\\
           &\hspace{0.5 cm}-4e^{-\frac{t\alpha_+}{\beta}}\beta^2\Bigg]+\frac{r}{\alpha_-^2}\Bigg[4\alpha_-\Bigg(\frac{t_c^3\ \Big[1-e^{-\frac{t}{t_c}(1+rt_c)}\Big]}{1+rt_c}-\frac{t_m^3\ \Big[1-e^{-\frac{t}{t_m}(1+rt_m)}\Big]}{1+rt_m}\Bigg)-\Bigg(\frac{t_c^4\ \Big[1-e^{-\frac{t}{t_c}(2+rt_c)}\Big]}{2+rt_c}\\
           &\hspace{0.5 cm}+\frac{t_m^4\ \Big[1-e^{-\frac{t}{t_m}(2+rt_m)}\Big]}{2+rt_m}\Bigg) +\frac{4 \beta^3\Big[1-e^{-\frac{t}{\beta}(r\beta+\alpha_+)}\Big]}{\alpha_+\big(r\beta+\alpha_+\big)}\Bigg]\Bigg\},
        \end{split}
    \end{equation}
\end{widetext}
where $\alpha_+=t_c+t_m$, $\alpha_-=t_c-t_m$ and $\beta=t_c t_m$.
In the limit $r \to 0$, $\langle\Delta x^2 (t) \rangle_r$ becomes equal to $\langle\Delta x^2(t) \rangle_0$ [Eq. \eqref{eq.ud_v_msd0}]. For very high resetting rate, $i.e., $ in the limit $r \to \infty$, $\langle \Delta x^2(t)\rangle_r$ approaches zero. In the limit $t \to 0$, $\langle\Delta x^2 (t) \rangle_r$ becomes
\begin{equation}\label{eq:ud_v_msdr_tsmall}
    \underset{t \to 0}{\lim }\langle\Delta x^2 (t) \rangle_r=v_0^2 t^2+\left[\frac{2k_BT-3t_m\gamma_fv_0^2}{3t_m^2\gamma_f}-\frac{2r}{3}v_0^2\right]t^3,
\end{equation}
indicating the initial ballistic motion of the particle. As time progresses, due to the resetting mechanism, the system is constantly reset to the initial state and leads the system to reach at a steady-state. 

In the time asymptotic limit (i.e., $t\to \infty$ limit), the $\langle\Delta x^2 (t) \rangle_r$ saturates to the steady state value ($\langle\Delta x^2(t) \rangle_r^{st}$) and it is given by
\begin{widetext}
\begin{equation}\label{eq.ud_v_msdr_st}
    \begin{split}
        \langle\Delta x^2(t) \rangle_r^{st} = \frac{4k_BT + 2rt_m^2v_0^2\gamma_f}{r(1+rt_m)(2+rt_m)\gamma_f}
        +\frac{4f_0^2t_c(2t_c+2t_m+3rt_ct_m)}{r(1+rt_c)(2+rt_c)(1+rt_m)(2+rt_m)(t_c+t_m+rt_ct_m)\gamma_f^2}.
    \end{split}
\end{equation}
\end{widetext}
In both the limits of $t_c \to 0$ and $t_c \to \infty$, the second term of Eq. \eqref{eq.ud_v_msdr_st} vanishes and $\langle\Delta x^2 (t) \rangle_r^{st}$ becomes $\frac{4 k_BT + 2rt_m^2v_0^2\gamma_f}{r(1+rt_m)(2+rt_m)\gamma_f}$, which is independent of $t_c$. 

\begin{figure*}[t]
    \centering
    \includegraphics[width=\textwidth]{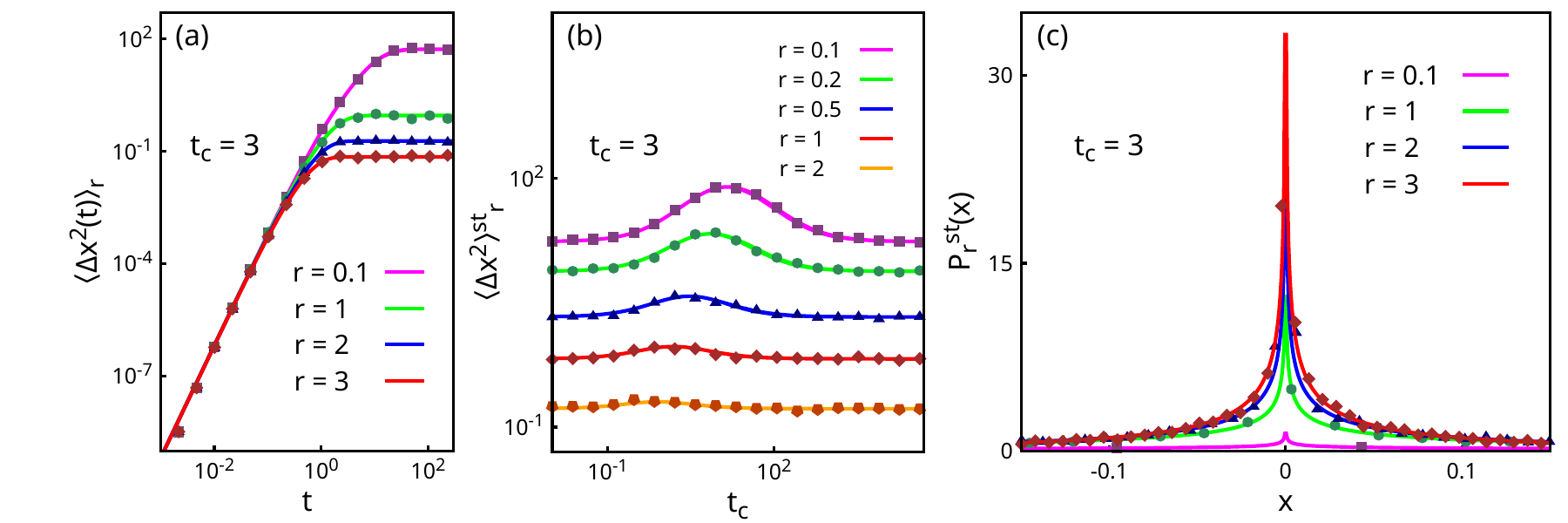}
    \caption{\justifying $\langle\Delta x^2(t)\rangle_r$ of an underdamped particle as a function of $t$ in $(a)$, $\langle\Delta x^2(t)\rangle_r^{st}$ as a function of $t_{c}$ in $(b)$, and $P_r^{st}(x)$ as a function of $x$ in $(c)$, are plotted respectively for different values of $r$. The other common parameters are $\gamma_f= f_0=t_m=k_BT=1.0$, and $v_0=0.0$. The solid lines represent analytical results and the markers represent the results from simulation.}
    \label{fig: ud_v_all}
\end{figure*}

Next, we have plotted $\langle\Delta x^2(t)\rangle_r$ as a function of $t$, $\langle\Delta x^2\rangle_r^{st}$ as a function of $t_c$, and $P_r^{st}(x)$ as a function of $x$ for different values of $r$ along with the simulation results in Figs. \ref{fig: ud_v_all} (a), (b), $\&$ (c), respectively. In the short-time regime, the MSD evolves with time as per Eq. \eqref{eq:ud_v_msdr_tsmall}. However, to make the calculation simple, we have set $v_0=0$ throughout our calculation. Therefore, in Fig. \ref{fig: ud_v_all}(a), $\langle\Delta x^2(t)\rangle_r$ grows as proportional to $t^3$ in the short-time regime and saturates to the steady-state value in the long-time limit. The effect of resetting is significant only in the steady state. The steady-state MSD is suppressed with an increase in the value of $r$. At the same time, MSD approaches steady state faster. For a fixed value of $r$, $\langle\Delta x^2\rangle_r^{st}$ shows a nonmonotonic dependence on $t_{c}$ [Fig. \ref{fig: ud_v_all}(b)]. It is same and independent of $t_{c}$ for both lower as well as larger values of $t_{c}$. In these limits, the impact of activity is negligible and the system behaves like a passive Brownian system. For intermediate values of $t_c$, the steady-state MSD is higher than the passive case. This enhancement in steady-state MSD enables the particle to explore a larger region thereby increasing the probability of reaching targets that are far away from the initial position. In the intermediate regimes of $t_c$, $\langle \Delta x^2\rangle_r^{st}$ increases with $t_{c}$, reaches a maximum value and then decreases with $t_{c}$. However, this non-monotonic impact of MSD on $t_{c}$ gets weakened for larger values of $r$, reflecting the dominant effect of resetting over the influence of activity in the dynamics. Now, substituting $P_0(x,t)$ for an underdamped particle in Eq. \eqref{eq.renewal_pdf}, we have numerically evaluated the $P_r(x,t)$ and plotted the steady-state distribution $P_r^{st}(x)$ as a function of $x$ for different values of $r$ in Fig. \ref{fig: ud_v_all}(c). The distribution gets suppressed with increase in $r$ value, which complements with the reduction of steady state MSD with $r$ as in Fig. \ref{fig: ud_v_all}(a).

\subsection{In a viscoelastic medium}
In this section, we analyze the dynamical behaviour of the particle while suspended in a viscoelastic environment characterized by the presence of a finite memory. 
The dynamics of the particle [Eq. \eqref{eq:lang_eqn_wo_r}] can be described by introducing a time dependent memory kernel in the viscous drag and follows the generalized Langevin equation (GLE) of motion \cite{goychuk2012viscoelastic,sevilla2019generalized,muhsin2021orbital,muhsin2023memory,muhsin2025active,adersh2025transition,adersh2025active}, given by

\begin{equation}\label{eq.ud_gle}
    m\dot v=-\int_0^t\gamma(t-t')\dot x(t')dt'+f_a(t)+\eta(t).
\end{equation}
Here, the first term in the right hand side of Eq. \eqref{eq.ud_gle} represents the viscoelastic drag force, with $\gamma(t-t')$ as the viscous kernel or memory kernel. In particular, we have considered the viscoelastic bath to be Jeffrey's fluid and hence the $\gamma(t-t')$ takes the form \cite{raikher2010theory,raikher2013brownian,das2023enhanced,biswas2025resetting,muhsin2025active}
\begin{equation}\label{eq.jeffreys_model}
    \gamma(t-t')=2\gamma_{f}\delta(t-t')+\frac{\gamma_s}{t_s}\exp{\left[ -\frac{(t-t')}{t_s} \right]}.
\end{equation}

Here, $\gamma_s$ denotes the strength of viscoelastic drag experienced by the particle and $t_s$ denotes the viscoelastic relaxation timescale. 
$\eta(t)$ is the thermal noise with following statistical properties:
\begin{equation}\label{eq.od_ve_noises_stat_prop}
\begin{split}
    \langle \eta(t)\rangle=0 \quad \quad ; \quad \quad \ \ \ \quad \langle \eta(t) \eta(s) \rangle =k_BT\gamma(t-s). \\
\end{split}
\end{equation}
In the overdamped limit, Eq. \eqref{eq.ud_gle} takes the form
\begin{equation}\label{eq.od_gle}
    \gamma_{f}\dot x(t)=-\int_0^t\frac{\gamma_s}{t_s}\exp{\left[-\frac{(t-t')}{t_s} \right]}\dot x(t')dt'+f_a(t)+\eta(t).
\end{equation}
The thermal noise $\eta(t)$ can be represented as the sum of two independent Gaussian noises such that $\eta(t)=\eta_0(t)+\eta_1(t)$, with properties $\langle\eta_0(t)\rangle=\langle\eta_1(t)\rangle=0$, $\langle\eta_0(t)\eta_0(s)\rangle=2\gamma_fk_BT\delta(t-s)$, and $\langle\eta_1(t)\eta_1(s)\rangle=\frac{\gamma_s}{t_s}k_BTe^{-\left(\frac{\left|t-s\right|}{t_s} \right)}$.
Here, $\eta_1(t)$ can be modeled as an OU process whose integral form can be written as \cite{das2023enhanced}
\begin{equation}\label{eq.od_ve_eta1_int_form}
    \eta_1(t)=\frac{1}{t_s}\int_0^t\exp\left[{-\frac{(t-t')}{t_s}}\right] \phi(t')dt'.
\end{equation}
Here, $\phi(t)$ is a Gaussian white noise such that $\langle\phi(t)\rangle=0$ and $\langle\phi(t)\phi(s)\rangle=2\gamma_sk_BT\delta(t-s)$. In $\gamma_s \to 0$ limit, the dynamics follows motion in a viscous medium as discussed in the previous section. 
In order for solving the dynamics, we introduce an auxiliary variable $w(t)$ such that
\begin{equation}\label{eq.od_ve_w_exp}
    w(t)=\frac{1}{t_s}\int_0^t\exp{\left[-\frac{(t-t')}{t_s} \right]}\left[\frac{\gamma_s}{t_s} x(t')+\phi(t')\right]dt'.
\end{equation}

Now, the dynamics of the system can be described by the following sets of Markovian equations
\begin{equation*}\label{eq.od_ve_dyn_x}
    \dot x(t)=-\frac{\gamma_s}{\gamma_ft_s}x(t)+\frac{1}{\gamma_f}w(t)+\frac{1}{\gamma_f}f_a(t)+\frac{1}{\gamma_f}\eta_0(t),
\end{equation*}
\begin{equation*}\label{eq.od_ve_dyn_w}
    \dot w(t)=\frac{\gamma_s}{t_s^2}x(t)-\frac{1}{t_s}w(t)+\frac{1}{t_s}\phi(t),
\end{equation*}
and
\begin{equation}\label{eq.od_ve_dyn_fa}
    \dot f_a(t)=-\frac{1}{t_c}f_a(t)+f_0\sqrt{\frac{2}{t_c}}\zeta(t).
\end{equation}
In the absence of resetting, the probability distribution $P_0(x,w,f_a,t)$ evolves according to the following FPE
\begin{widetext}
\begin{equation}\label{eq.od_ve_fpe}
\begin{split}
    \frac{\partial P_0}{\partial t}&=\frac{\partial}{\partial x}\bigg(\frac{\gamma_s x}{\gamma_ft_s}-\frac{w}{\gamma_f}-\frac{f_a}{\gamma_f}+\frac{k_BT}{\gamma_f}\frac{\partial}{\partial x} \bigg)P_0+\frac{\partial}{\partial w}\bigg(-\frac{\gamma_s x}{t_s^2}+\frac{w}{t_s}+\frac{\gamma_s k_BT}{t_s^2}\frac{\partial}{\partial w}\bigg)P_0+\frac{\partial}{\partial f_a}\bigg( \frac{f_a}{t_c}+\frac{f_0^2}{t_c}\frac{\partial}{\partial f_a} \bigg)P_0. 
\end{split}
\end{equation}
\end{widetext}

First, we calculate the MSD of the particle using the Laplace transform method.
The dynamics [Eq. \eqref{eq.od_gle}] in the Laplace domain can be obtained as 
\begin{equation}\label{eq.od_gle_lap}
    \left[ \gamma_f +\frac{\gamma_s}{1+st_s} \right] \left[ s \tilde{x}(s) -x_0\right] = \tilde{f_a}(s) +\tilde{\eta}(s),
\end{equation}
where $\tilde{x}(s)=\int_0^\infty e^{-st}x(t)dt,\ \tilde{\eta}(s)=\int_0^\infty e^{-st}\eta(t)dt$ and $\tilde{f_a}(s)=\int_0^\infty e^{-st}f_a(t)dt$ are Laplace-transformed quantities. $x_0$ is the initial position of the particle. Now, the position of the particle in the Laplace domain can be obtained as 
\begin{equation}\label{eq.od_ve_x_lap}
    \tilde{x}(s)=\frac{x_0}{s}+\frac{1+st_s}{s\left[ s+\frac{\gamma_f+\gamma_s}{\gamma_f t_s} \right]\gamma_f t_s} \left[  \tilde{f_a}(s)+\tilde{\eta}(s)  \right]. 
\end{equation}
By taking the inverse Laplace transform of Eq. \eqref{eq.od_ve_x_lap}, we get the position of the particle as 

\begin{widetext}
    \begin{equation}\label{eq.od_ve_x_exp}
    \begin{split}
     x(t)=x_0+\frac{1}{\left( \gamma_f+\gamma_s \right)}\int_0^t\Bigg[ 1+\frac{\gamma_s}{\gamma_f} e^{-\left( \frac{\gamma_f+\gamma_s}{\gamma_f} \right)\frac{\left( t-t' \right)}{t_s}} \Bigg] \Bigg[  f_a(t')+\eta(t')  \Bigg] dt'. 
    \end{split} 
    \end{equation}
Throughout the calculation we set $x_0=0$. Hence the displacement $\Delta x(t)=x(t)$. Using Eq. \eqref{eq.msd0_formula}, we obtain the MSD of an underdamped particle without resetting as
    \begin{equation}\label{eq.od_ve_msd0}
    \begin{split}
        \langle\Delta x^2(t) \rangle_0&=\frac{1}{g^2}\Bigg\{2k_BT\Big[gt+\gamma_s t_s\Big(1-e^{-\frac{gt}{\gamma_f t_s}}\Big)\Big]+f_0^2t_c\Bigg[2t+\frac{1}{g j_+ j_-^2}\Bigg(4 g^2 j_+ j_- t_c(t_c-t_s)e^{\frac{-t}{t_c}}\\
        &\hspace{0.5 cm}-g^3 j_+ t_c(t_c-t_s)^2e^{\frac{-2t}{t_c}}+4j_+ j_- \gamma_f \gamma_s t_s^2 e^{-\frac{g t}{\gamma_f t_s}}-4 g^2 \gamma_f \gamma_s t_c t_s^2 (t_c-t_s) e^{-t\Big(\frac{1}{t_c}+\frac{g}{\gamma_f t_s}\Big)}-j_+ \gamma_f \gamma_s^2 t_s^3 e^{\frac{-2gt}{\gamma_f t_s}}\\
        &\hspace{0.5 cm}-j_-^2\Big[3 g^2 t_c^2+g t_c t_s \Big(3 \gamma_f -2 \gamma_s\Big)-\gamma_s t_s^2\Big(4 \gamma_f +\gamma_s\Big)\Big] \Bigg)\Bigg]\Bigg\}.          
    \end{split}
    \end{equation}
\end{widetext}
Here, $g=\gamma_f + \gamma_s$, $j_+=gt_c+\gamma_ft_s$ and $j_-=gt_c-\gamma_ft_s$. In the limit $t \to 0$, the $\langle\Delta x^2 (t) \rangle_0$ in Eq. \eqref{eq.od_ve_msd0} approaches $\frac{2k_BT}{\gamma_f}t$, which is proportional to $t$. Therefore, at short time scales, the viscous force dominates and the particle exhibits a diffusive motion with an effective diffusion coefficient $\frac{k_BT}{\gamma_f}$. In the limit $t \to \infty$, $\langle\Delta x^2(t) \rangle_0$ becomes
\begin{equation}\label{od_ve_msd0_t_long}
    \underset{t \to \infty}{\lim}\langle\Delta x^2(t) \rangle _0=\left\{\frac{2k_BT}{\gamma_f+\gamma_s}+\frac{f_0^2t_c}{\left( \gamma_f+\gamma_s\right) ^2}\right\}t.
\end{equation}
Eq. \eqref{od_ve_msd0_t_long} shows that at longer times, the particle is again diffusive but the rate of diffusion depends upon the strength of the viscoelastic drag and the activity timescale in addition to the viscous coefficient. However, $\langle\Delta x^2 (t)\rangle_0$ becomes independent of time and remains constant between the short time and long time diffusion processes, 
giving rise to an intermediate plateau region as reflected from Eq. \eqref{eq.od_ve_msd0}. 

When stochastic resetting is introduced, the dynamics of the particle is reset to the initial state at stochastic intervals with a constant resetting rate $r$. The reset mechanism can be described as

\begin{equation}\label{od_ve_r_dynamics}
    \textbf{X}(t+dt)= \begin{cases}
        \textbf{X}(0) \quad \quad \quad \quad \quad \ \ $with probability$ \ rdt \\
        Eq.\hspace{0.1cm}\eqref{eq.od_ve_dyn_fa} \ \quad \quad \quad $with probability$\ 1 - rdt \\ 
    \end{cases}    
\end{equation}

Here, $\textbf{X}(t)=(x(t),w(t),f_a(t))^T$ and the initial configuration is $\textbf{X}(0)=(0,0,f_{a0})^T$. 
Substituting $\langle\Delta x^2(t)\rangle_0$ in Eq. \eqref{eq.msd_renewal_form}, we have analytically calculated the MSD with resetting $\langle\Delta x^2(t) \rangle _r$. The exact expression of $\langle\Delta x^2(t) \rangle_r$ is given in Appendix \ref{app.ve_msdr}. In the limit $t \to 0$, $\langle\Delta x^2(t) \rangle _r$ approaches $\frac{2k_BT}{\gamma_f}t$. When we expanded $\langle\Delta x^2(t) \rangle_r$ in powers of $t$, we obtain a timescale $t_{short}  =\frac{2 \gamma_f t_s }{r t_s \gamma_f +\gamma_s}$, upto which all the higher order terms in $t$ are negligible and $\langle\Delta x^2 (t) \rangle_r \approx \frac{2k_BT}{\gamma_f}t$. When $t \gg t_{short}$ and in the limit $t \to \infty$, the exponential terms of $\langle \Delta x^2(t)\rangle_r$ in Eq. \eqref{od_ve_msdr_full_exp} can be eliminated. In this limit,  $\langle \Delta x^2(t)\rangle_r$ can be approximated to the steady-state MSD $\langle\Delta x^2(t)\rangle_r^{st}$ and it is given by
\begin{widetext}
    \begin{equation}\label{eq.od_ve_msdr_st}
    \begin{split}
        \langle\Delta x^2 (t) \rangle_r^{st} &=\frac{2}{r \big[\gamma_f (1+r t_s)+\gamma_s\big]}\Bigg\{k_BT(1+ rt_s)\\
        &\hspace{0.5 cm}+\frac{2 f_0^2 t_c\Big\{\gamma_f (1+rt_s) (2+rt_s)\big[t_s + t_c(1+rt_s)\big]+\gamma_s \Big[t_c[2+rt_s(2 +rt_s)]+rt_s^2 \Big]\Big\}}{(1+rt_c)(2+rt_c)\big\{ \gamma_f t_s + t_c \big[\gamma_f(1 +r t_s) + \gamma_s\big] \big\}\big[ (2+rt_s)\gamma_f + 2\gamma_s \big] } \Bigg\}. 
    \end{split}
\end{equation}
\end{widetext}
From Eq. \eqref{eq.od_ve_msdr_st}, it is confirmed that the system with resetting reaches at a steady state. This is because unlike the case of the system without resetting where the steady state is diffusive [Eq. \eqref{od_ve_msd0_t_long}], $\langle \Delta x^2(t) \rangle_r^{st}$ is independent of time and is entirely due to the effect resetting mechanism. 
In the limit $t \gg t_{short}$, when $r$ is small, $e^{-rt}$ cannot be eliminated from $\langle\Delta x^2(t) \rangle_r$. Hence, for small $r$ and for $t\gg t_{short}$, $e^{-rt}\approx 1-rt$ and $\langle \Delta x^2(t) \rangle_r$ approximately becomes $\langle \Delta x^2(t)\rangle_{int}$ and it is given by
\begin{widetext}
\begin{equation}\label{od_ve_msd0_int_and_tlong}
\begin{split}
    \langle\Delta x^2(t) \rangle_{int} &\approx \Bigg\{ \langle x^2(t) \rangle_r^{st}-\frac{2\big[f_0^2 t_c+k_BT( \gamma_f +\gamma_s)\big]}{r \big(\gamma_f + \gamma_s\big)^2}\Bigg\} + \frac{2\big[f_0^2 t_c +k_BT(\gamma_f +\gamma_s)\big]}{ \big(\gamma_f + \gamma_s\big)^2}t\\ 
    &\approx \Bigg[\langle x^2(t) \rangle_r^{st}-\frac{2\big[f_0^2 t_c+k_BT (\gamma_f +\gamma_s)\big]}{r \big(\gamma_f + \gamma_s\big)^2} \Bigg] \Bigg[1 + \frac{t}{t_{pl}} \Bigg],
\end{split}
\end{equation}
where
\begin{equation}\label{eq. od_ve_msdr_tint}
\begin{split}
     t_{pl}&= \frac{(\gamma_f+\gamma_s)^2}{r\big[f_0^2t_c+k_BT(\gamma_f+\gamma_s)\big]\big[\gamma_f(1+rt_s)+\gamma_s\big]}\Bigg\{k_BT(1+rt_s)\\
     &\hspace{0.5 cm}+\frac{2 f_0^2 t_c\big\{ (1+rt_s)(2+rt_s)(t_c+t_s+rt_ct_s)\gamma_f +rt_s^2\gamma_s+ t_c \big[ 2+rt_s(2+rt_s)  \big]\gamma_s\big\}}{(1+rt_c)(2+rt_c)\big[(2+rt_s)\gamma_f +2\gamma_s \big]\big\{ t_s \gamma_f +t_c[\gamma_f(1+rt_s)+\gamma_s] \big\}}\Bigg\}-\frac{1}{r}.
\end{split}   
\end{equation}
When $t \ll t_{pl}$, the second term in Eq. \eqref{od_ve_msd0_int_and_tlong} is negligible and $\langle\Delta x^2 (t) \rangle_{int}$ can be approximated to $\langle \Delta x^2(t) \rangle_{pl}$ and it is given by 
\begin{equation}\label{eq.od_ve_msdr_int}
    \begin{split}
        \langle\Delta x^2(t) \rangle_{pl}= \langle \Delta x^2(t) \rangle_r^{st}-\frac{2\big[f_0^2t_c+ k_BT (\gamma_f +\gamma_s)\big]}{r \big(\gamma_f + \gamma_s\big)^2}. 
    \end{split}  
\end{equation}
\end{widetext}
From Eq. \eqref{eq.od_ve_msdr_int}, it is confirmed that $\langle\Delta x^2(t) \rangle_{pl}$ corresponds to the time-independent intermediate plateau. When $t\gg t_{pl}$, the second term in Eq. \eqref{od_ve_msd0_int_and_tlong} becomes dominant and the MSD grows linearly with time. As time progresses, the effect of resetting becomes more significant and it drives the system to reach at steady-state. 


\begin{figure*}[!ht]
    \centering
    \includegraphics[width=0.7\textwidth]{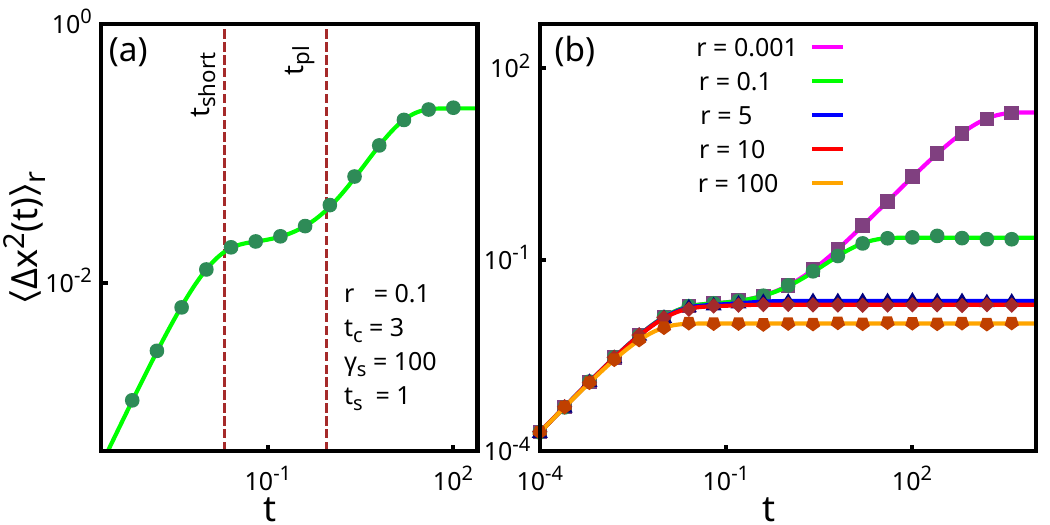}
    \caption{\justifying$\langle\Delta x^2(t) \rangle_r$ of an overdamped particle (in viscoelastic medium) as a function of $t$ for $r=0.1$ in $(a)$ with the dotted vertical lines representing the timescales $t_{short}$ and $t_{pl}$, and for different values of $r$ in $(b)$. The other common parameters are $\gamma_f=f_0=t_s=k_BT=1.0$, $\gamma_s=100.0$, and $t_c=3.0$. The solid lines represent analytical results and the markers represent the results from simulation.}
    \label{fig:od_ve_msdr_r}
\end{figure*}

In Fig. \ref{fig:od_ve_msdr_r}, we have plotted $\langle\Delta x^2(t)\rangle_r$ as a function of $t$ for $r=0.1$ in Fig. \ref{fig:od_ve_msdr_r}(a) with vertical dotted lines representing the timescales $t_{short}$ and $t_{pl}$ and for different values of $r$ in Fig. \ref{fig:od_ve_msdr_r}(b) . For a fixed value of $r$ and when $r$ is small enough, i.e., for $r \ll 1$, $\langle\Delta x^2(t)\rangle_r$ initially increases linearly with $t$. This linear time dependence of $\langle\Delta x^2(t)\rangle_r$ occurs up to time $t_{short}$, then displays an intermediate time-independent plateau for a time interval of $t_{pl}$. After $t_{pl}$, $\langle \Delta x^2(t) \rangle_r$ increases linearly again with $t$ and finally saturates to the steady-state value in the time asymptotic limit [Fig. \ref{fig:od_ve_msdr_r}(a)]. With an increase in the value of $r$, $\langle \Delta x^2(t) \rangle_r$ approaches the steady state value faster and the intermediate plateau disappears [see Fig. \ref{fig:od_ve_msdr_r}(b)]. This is because, with an increase in the value of $r$, the contribution of the second term of Eq. \eqref{eq.od_ve_msdr_int} to the $\langle \Delta x^2(t) \rangle$ decreases, as a result of which the steady state value is approached faster and for very large value of $r$, $\langle \Delta x^2(t) \rangle_{pl}$ becomes $\langle \Delta x^2 (t) \rangle_r^{st}$.


Similarly, for short persistence of activity in the medium, i.e., in the limit $t_c \to 0$, the second term of Eq. \eqref{eq.od_ve_msdr_st} vanishes and $\langle\Delta x^2 (t) \rangle_r ^{st}$ becomes 
\begin{equation}\label{eq.od_ve_msdr_tc_small}
  \underset{t_c \to 0}{\lim}\langle\Delta x^2 (t) \rangle _r ^{st} = \frac{2 k_BT\left( 1+rt_s \right)}{r\big[\left( 1+rt_s \right)\gamma_f +\gamma_s\big]},
\end{equation}
which is the steady state MSD of a Brownian particle in a viscoelastic bath under stochastic resetting \cite{biswas2025resetting}. In both limits of $t_c \to 0$ and $\gamma_s \to 0$, it becomes $\frac{2 k_BT}{r\gamma_f}$, which is same for a Brownian particle subjected to stochastic resetting \cite{gupta2022stochastic}. For a large value of $t_c$, the strength of the correlation of active force decreases. Hence, in the limit $t_c \to \infty$ the second term in Eq. \eqref{eq.od_ve_msdr_st} vanishes and $\langle\Delta x^2 (t) \rangle _r ^{st}$ becomes same as in Eq. \eqref{eq.od_ve_msdr_tc_small}.
\begin{figure}[h]
    \centering
    \includegraphics[width=\linewidth]{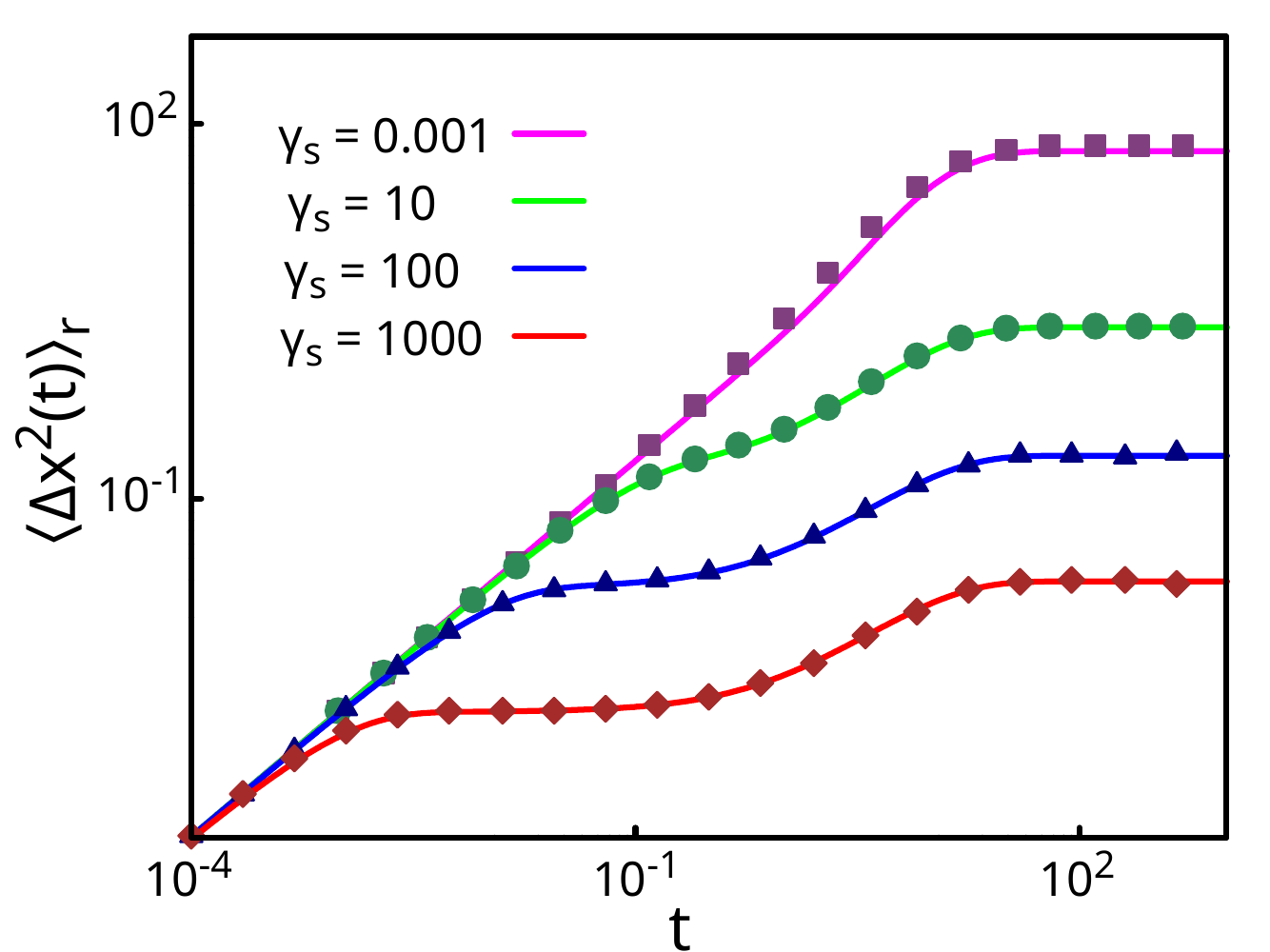}
    \caption{\justifying$\langle\Delta x^2(t) \rangle_r$ of an overdamped particle (in viscoelastic medium) as a function of $t$ for different values of $\gamma_s$. The other common parameters are $\gamma_f=f_0=t_s=k_BT=1.0$, $r=0.1$, and $t_c=3.0$. The solid lines represent analytical results and the markers represent the results from simulation.}
    \label{fig:od_ve_msdr_gs}
\end{figure}
The intermediate plateau in $\langle \Delta x^2(t)\rangle_r$ appears due to the complex interplay of activity, viscous and viscoelastic drag forces. In $\gamma_s \to 0$ limit, $\langle\Delta x^2(t) \rangle_r$ approaches the MSD of an overdamped particle in viscous medium as in Eq. \eqref{eq.od_msdr_full_exp} with no intermediate plateau. Moreover, in the limit $\gamma_s \to \infty$, $\langle\Delta x^2(t) \rangle_r$ approaches zero value. 
In the Fig. \ref{fig:od_ve_msdr_gs}, we have plotted $\langle\Delta x^2(t) \rangle_r$ as a function of $t$ for different values of $\gamma_s$. For a fixed value of $\gamma_s$, $\langle \Delta x^2(t)\rangle_r$ initially varies linearly with $t$ for a short time regime ($t_{short}$), then in the intermediate time regime ($t_{short}\ll t \ll t_{pl}$), $\langle\Delta x^2(t)\rangle_r$ becomes time-independent giving rise to a saturated value or plateau. After a time $t \gg t_{pl}$, it again varies linearly with $t$ and finally saturates to the steady-state value. The intermediate time plateau regime has a significant impact on $\gamma_s$. For very small value of $\gamma_s$, the plateau regime is not prominent and becomes significant with increase in the value of $\gamma_{s}$. With increase in $\gamma_s$ value, the viscoelastic memory effects becomes stronger and the time interval of $t_{short}$ decreases. As a result, the intermediate time plateau starts appearing much earlier, leading to the increase in the length of the plateau regime. At the same time, the steady-state MSD ($\langle \Delta x^2(t) \rangle_r^{st}$) gets suppressed, reflecting the restriction of motion of the particle under the influence of viscoelastic drag force.  

\begin{figure}[h]
    \centering
    \includegraphics[width=\linewidth]{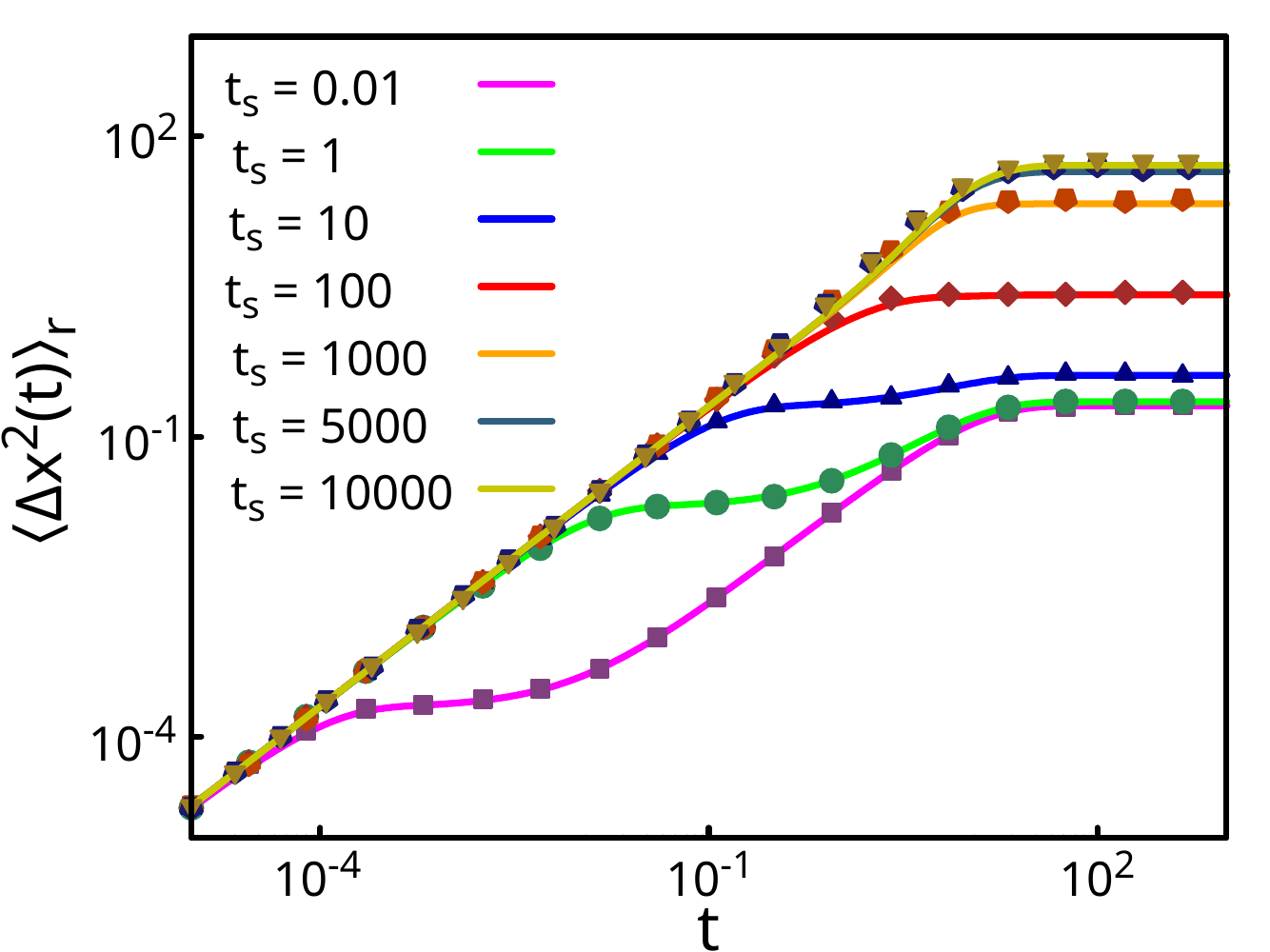}
    \caption{\justifying$\langle\Delta x^2(t) \rangle_r$ of an overdamped particle (in viscoelastic medium) as a function of $t$ for different values of $t_s$. The other common parameters are $\gamma_f=f_0=k_BT=1.0$, $\gamma_s=100.0$, $r=0.1$, and $t_c=3.0$. The solid lines represent analytical results and the markers represent the results from simulation.}
    \label{fig:od_ve_msdr_ts}
\end{figure}

In Fig. \ref{fig:od_ve_msdr_ts} we have plotted $\langle\Delta x^2 (t) \rangle_r$ as a function of $t$ for different values of $t_s$. For a fixed value of $t_s$, the $\langle\Delta x^2 (t) \rangle_r$ initially increases linearly with $t$, saturates in the intermediate time regime, then grows again as a linear function of $t$. In the long time limit, it saturates to the steady-state value. From Fig. \ref{fig:od_ve_msdr_ts}, it is observed that for the lower and higher values of $t_{s}$, the steady-state value of $\langle\Delta x^2(t)\rangle_r$ does not get affected by $t_{s}$. In the limit $t_s \to 0$, $\langle\Delta x^2(t)\rangle_r^{st}$ becomes $\frac{2}{r(\gamma_f+\gamma_s)}\Big[ k_BT + \frac{2 f_0^2 t_c}{(1+rt_c)(2+rt_c)(\gamma_f+\gamma_s)} \Big]$. Similarly, in the limit $t_s \to \infty$,  $\langle\Delta x^2(t)\rangle_r^{st}$ becomes $\frac{2}{r \gamma_f}\Big[k_BT + \frac{2 f_0^2 t_c}{(1+rt_c)(2+rt_c)\gamma_f} \Big]$.   
From both of these limiting behaviors, it is confirmed that for very small $t_s$ and for very large $t_s$, $\langle \Delta x^2(t)\rangle_r^{st}$ becomes independent of $t_s$ and the variation of the steady-state MSD is significant only for intermediate values of $t_s$. With increase in the value of $t_s$, the $t_{short}$ shifts towards right and the short time regime increases. As a result, the intermediate plateau starts appearing later and the width of the intermediate plateau decreases, as reflected from Fig. \ref{fig:od_ve_msdr_ts}. 


The probability distribution without resetting $P_{0}(x,w,f_a,t)$ evolves according to Eq. \eqref{eq.od_ve_fpe} which is a linear differential equation and hence $P_0(x,w,f_a,t)$ takes a Gaussian form. Integrating $P_0(x,w,f_a,t)$ over all other variables, we get $P_0(x,t)$ as the same form of Eq. \eqref{eq.od_v_p0_exp}. Now, substituting $P_0(x,t)$ in Eq. \ref{eq.renewal_pdf}, we get the probability distribution with resetting $P_r(x,t)$. 
 \begin{figure}[h]
    \centering
    \includegraphics[width=\linewidth]{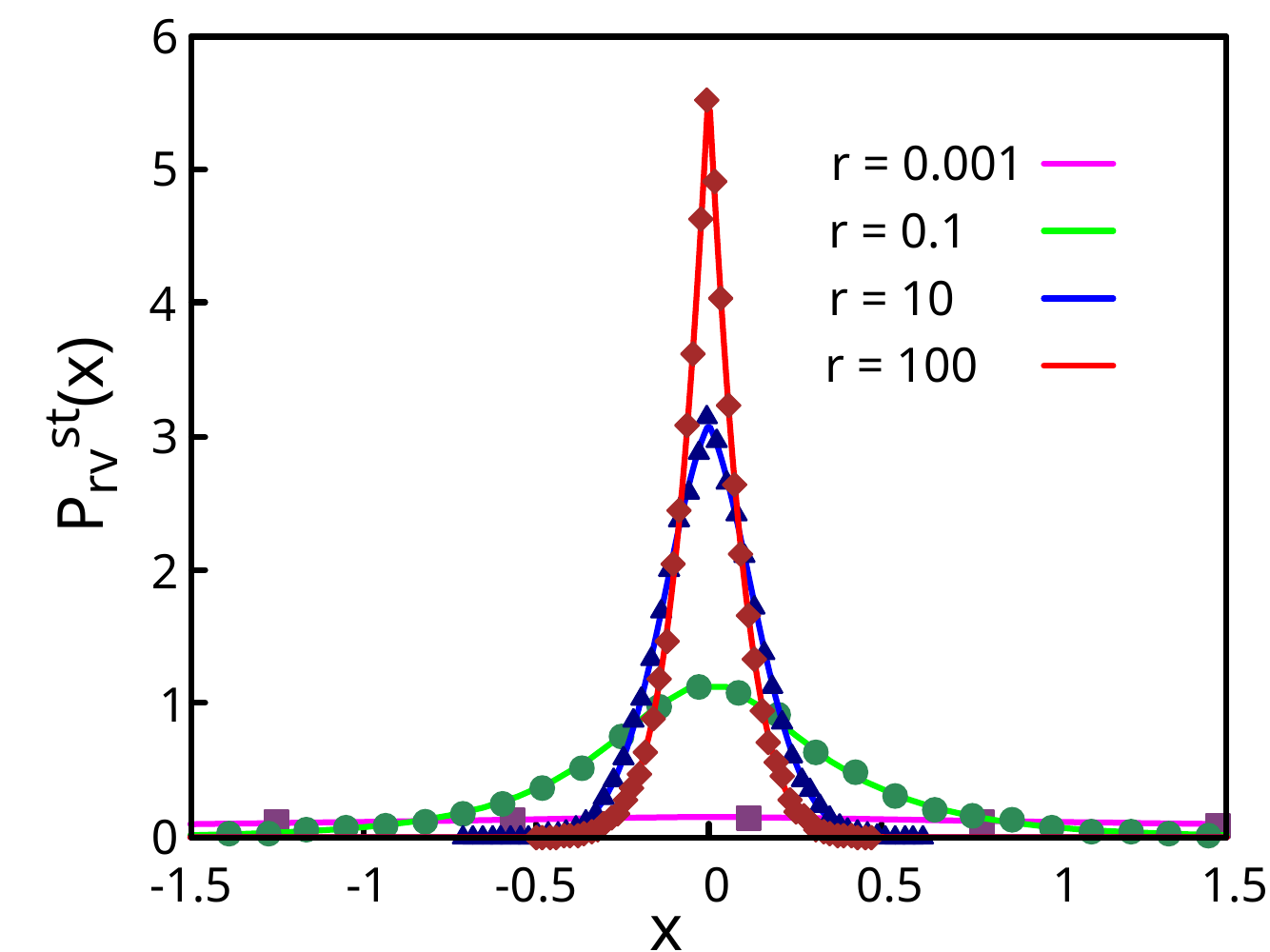}
    \caption{\justifying$P_{rv}^{st}(x)$ as a function of $x$ for different values of $r$. The common parameters are $\gamma_f=f_0=t_s=k_BT=1.0$, $\gamma_s=100.0$, and $t_c=3.0$. The solid lines represent analytical results and the markers represent the results from simulation.}
    \label{fig:od_ve_pr_r}
\end{figure}
We denote the steady-state probability distribution with resetting in viscoelastic environment as $P_{rv}^{st}(x)$. In Fig. \ref{fig:od_ve_pr_r}, we have plotted $P_{rv}^{st}(x)$ along with the simulated curves as a function of $x$ for different values of $r$. For a fixed value of $r$, the distribution remains non-Gaussian, with an exponential form centered at the origin. 
However, with an increase in the value of $r$, the resetting mechanism becomes dominant over all other involved processes, as a result the peak of the distribution becomes sharper, implying the enhancement of the probability of finding the particle at the resetting point. This behavior is consistent with the suppression of steady-state MSD with $r$ as in Fig. \ref{fig:od_ve_msdr_r}.
 
\begin{figure}[h]
    \centering
    \includegraphics[width=\linewidth]{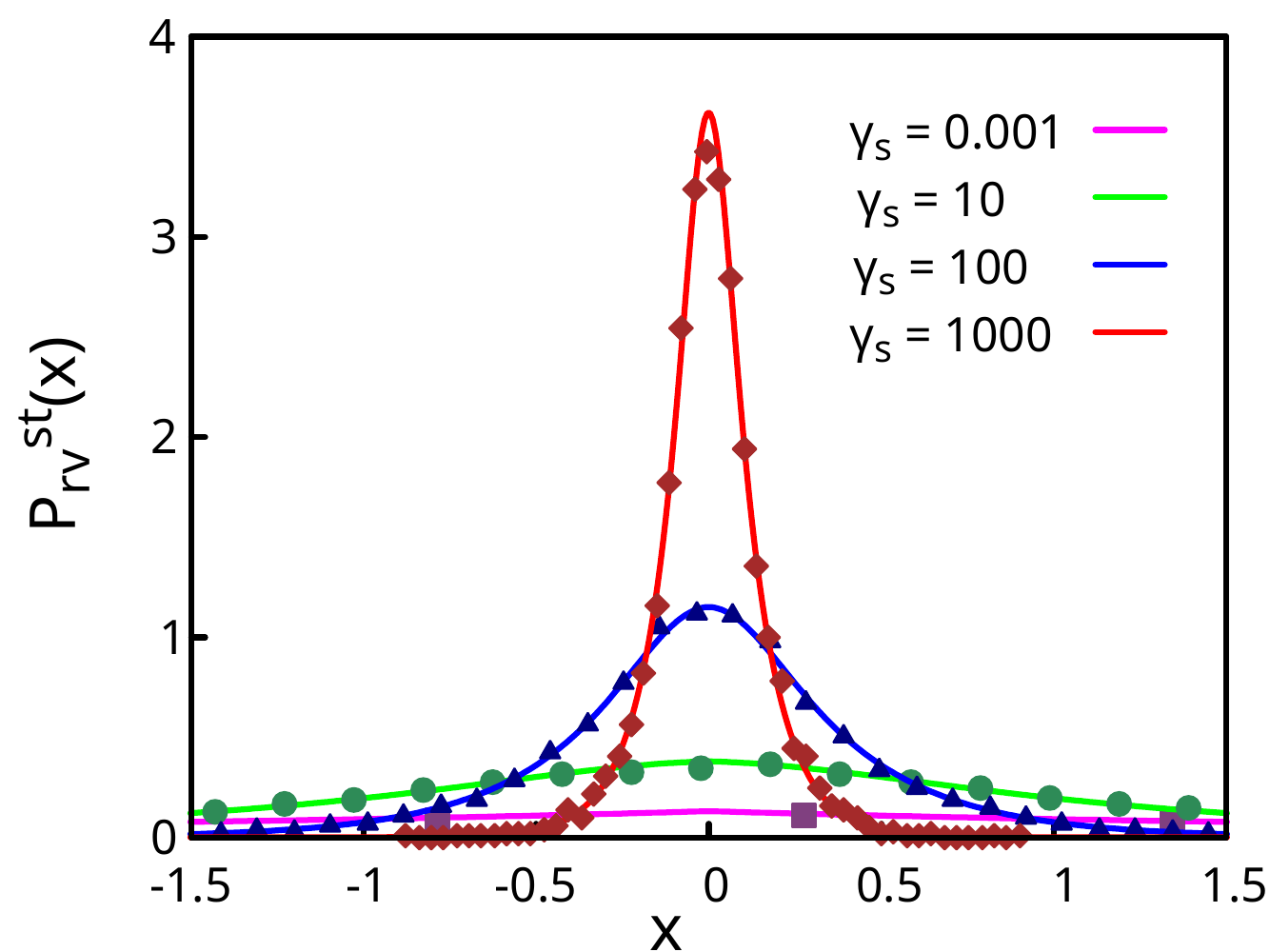}
    \caption{\justifying$P_{rv}^{st}(x)$ as a function of $x$ for different values of $\gamma_s$. The other common parameters are $\gamma_f=f_0=t_s=k_BT=1.0$, $r=0.1$, and $t_c=3.0$. The solid lines represent analytical results and the markers represent the results from simulation.}
    \label{fig:od_ve_pr_gs}
\end{figure}
In Fig. \ref{fig:od_ve_pr_gs}, we have plotted $P_{rv}^{st}(x)$ along with the simulated curves, as a function of $x$ for different values of $\gamma_s$. For a fixed value of $\gamma_s$, the distribution is exponentially centered at $x_0=0$ (resetting point). 
As the value of $\gamma_s$ increases, the strength of the viscoelastic drag increases, restricting the movement of the particle away from the origin. Hence, the distribution becomes narrower and the peak of the distribution increases with increase in $\gamma_s$ value, which complements the reduction of the steady-state MSD with increase in $\gamma_{s}$ as in Fig. \ref{fig:od_ve_msdr_gs}. For a very large value of $\gamma_s$, the peak becomes less pronounced. This is due to the stronger memory effects that weaken the localization induced by the resetting mechanism.    

\section{SUMMARY}\label{sec:summary}
In summary, we have investigated the dynamics of a free active Ornstein-Uhlenbeck particle in a viscous and viscoelastic environment subjected to stochastic resetting. We exactly explore the transport behavior using both the analytical method and numerical simulation. To understand the combined impact of activity and resetting on the diffusive behavior of the particle, we analytically computed the MSD and position probability distribution function. 
using a renewal approach, while suspended in a viscous as well as viscoelastic medium. 
From the MSD calculation, it is observed that the effect of resetting and the influence of activity are significant only in the steady state. The steady-state MSD is suppressed with frequent resetting and shows a nonmonotonic impact on the persistence duration of the activity. 
For a fixed resetting rate, for very short and long persistent durations of activity, the steady-state MSD is not affected by the persistent duration of activity. 
However, for an intermediate range of the duration of activity, the steady-state MSD increases with the activity time, shows a maximum, then decreases, and saturates to a constant value for a very large value of activity time. The enhancement of steady-state MSD for an intermediate range of activity timescales implies that the particle explores a larger region compared to its passive counterpart. However, this nonmonotonic behavior of MSD is significant only for a low resetting rate. Therefore, slow resetting and intermediate range of activity time are favorable conditions to reach a wide spread target. 

Moreover, while suspended in a viscoelastic environment, the MSD with resetting initially shows diffusive behavior, that is, it grows linearly with time, displays an intermediate time plateau, then again varies linearly with time, and finally saturates to a constant value in the long-time or time asymptotic limit. From both the analytical calculation and the simulation results, it is observed that the initial diffusive behavior depends on the time $t_{short}$. The increase in viscoelastic strength leads to stronger memory effects in the medium, which results in a decrease in time $t_{short}$ and an increase in the width of the intermediate plateau regime. Similarly, the slower relaxation time enhances the $t_{short}$, delays the onset of plateau, and for a very long relaxation time, the intermediate plateau disappears, resulting in the approach of steady state. However, the steady state MSD gets suppressed with increase in the viscoelastic strength, whereas long persistent duration of memory in the medium enhances the steady state MSD. Hence, low memory strength and longer persistence of memory are beneficial for a wide spread target search of an active particle in a viscoelastic medium. In addition, the frequent resetting of the dynamics leads the system towards steady state faster. 
Moreover, the influence of activity is significant only at the steady-state, and the steady-state MSD shows a nonmonotonic dependence on the activity timescale. Our simulation results are in good agreement with the analytical data.



We believe that our work provides a theoretical foundation for the design of efficient search strategies and targeted transport. 
It would be further interesting to investigate the transport using a different kernel, such as the multiexponential kernel \cite{klimek2025subdiffusion,klimek2025hierarchical}, power law kernel \cite{min2005observation,min2006kramers,sandev2015diffusion}, oscillatory kernel \cite{li2022non,reigada1999generalized,doerries2021correlation}. 
Another promising direction is to consider a non-Markovian resetting scheme \cite{pal2017first,radice2022diffusion} or a spatially distributed resetting rate \cite{evans2011diffusion_with_opt_resetting,pinsky2020diffusive} in dynamics and investigate the impact on the properties of steady state. 
\section{Acknowledgement}
MS acknowledges the computational facility, Department of Physics, University of Kerala, SERB-SURE grant (SUR/2022/000377), and CRG grant (CRG/2023/002026) from DST, Govt. of India for 
financial support.
\appendix 
\section{}{\label{app.correlation_matrix}}
In order to solve the FPE in Eq. \eqref{ud_fpe_wo_r} we describe the dynamics of the system [Eq. \eqref{eq:ud_dynamics}] in the matrix form
\begin{equation}\label{eq:drift_diff_matrix}
    \dot {\textbf{X}}=A\textbf{X}+B\boldsymbol{\eta'}.
\end{equation}
Here, the matrix $A$ represents drift term, the matrix $B$ represents the diffusion term and $\boldsymbol{\eta'}$ is noise vector whose components are $(0,\eta(t),\zeta(t))$.
Then, the FPE in Eq. \eqref{ud_fpe_wo_r} can be written as
\begin{equation}\label{eq;gen_fpe}
    \frac{\partial P_0}{\partial t}=\sum_{i,j=1}^3 \left\{-A_{ij}\frac{\partial}{\partial X_i}(X_j P_0)+\frac{1}{2}B_{ij}^2\frac{\partial^2 P_0}{\partial X_i \partial X_j}\right\}. 
\end{equation}
The solution to the linear FPE in \eqref{eq;gen_fpe} will be a Gaussian and it can be determined by knowing the first and second moments of $P_0(\textbf{X},t)$. To calculate the moments we use the correlation matrix formalism\cite{van1992stochastic,adersh2024inertial,muhsin2025active}.

The correlation matrix can be defined in the following way\vspace{-00.5 cm}
\begin{equation}\label{corr_matrix_form}
    [\Xi_{i,j}] = \langle X_i X_j\rangle - \langle X_i \rangle \langle X_j \rangle.
\end{equation}

As per\cite{van1992stochastic}, 
\begin{equation}
    \textbf{$\Xi$}(t)=\int_0^t e^{(t-t')A}Be^{(t-t')\tilde{A}} dt'.
\end{equation}

The corresponding probability distribution is given by,
\begin{equation}\label{ud_v_prob_x_xi_t}
\begin{split}
    P_0(\textbf{X},t) &= (2\pi)^{-1}(\det \Xi)^{-\frac{1}{2}}\\
    &\hspace{01 cm}\exp\Bigg\{{-\frac{1}{2}}\Big[\tilde{\textbf{X}}-\langle \tilde{\textbf{X}} \rangle\Big]\Xi^{-1}\Big[\textbf{X} - \langle \textbf{X} \rangle\Big]\Bigg\}
\end{split}
\end{equation}

Here, $\textbf{X}(t)=(x(t),v(t),f_a(t))^T$. Then, according to the definition in Eq. \eqref{corr_matrix_form}, $\Xi_{1,1}$ gives the variance in position, $\sigma=\langle\Delta x(t)^2 \rangle-\langle \Delta x(t)\rangle^2$. 
\section{}\label{app.ve_msdr}
The exact expression of the MSD of an overdamped particle while suspended in a viscoelastic environment and subjected to stochastic resetting is given by

\begin{widetext}
    \begin{equation}\label{od_ve_msdr_full_exp}
    \begin{split}
        \langle \Delta x^2(t) \rangle_r&=\frac{2}{r(g+r\gamma_f t_s)}\Bigg\{k_BT(1+rt_s)+\frac{2 f_0^2 t_c\Big\{(1+rt_s)(2+rt_s)(t_c+t_s+rt_ct_s)\gamma_f+r\gamma_st_s^2+\gamma_st_c\Big[2+rt_s(2+rt_s)\Big]\Big\}}{(1+rt_c)(2+rt_c)(gt_c+(1+rt_c)\gamma_ft_s)\Big[(2+rt_s)\gamma_f+2\gamma_s\Big]}\Bigg\}\\
        &+\frac{1}{g^2}r\Bigg\{-\frac{2k_BT\gamma_f^3\gamma_st_s^4e^{-t\big(r+\frac{g}{\gamma_ft_s}\big)}}{(g+r\gamma_ft_s)j_+j_-}+t_c\Bigg[\frac{2f_0^2t_c e^{-rt}}{r}-\frac{3f_0^2t_c^2g(t_c-t_s)e^{-\frac{t}{t_c}(1+rt_c)}}{j_-(1+rt_c)}+\frac{\Big[f_0 t_c g(t_c-t_s)\Big]^2e^{-\frac{t}{t_c}(2+rt_c)}}{j_-^2(2+rt_c)}\\
        &-\frac{f_0^2t_c^2g\gamma_f\gamma_st_s^2(j_--2\gamma_ft_s)(t_c-t_s)e^{-\frac{t}{\gamma_ft_ct_s}\big[gt_c+(1+rt_c)\gamma_ft_s\big]}}{gt_c+(1+rt_c)}-\frac{2\gamma_f\gamma_st_s^2 e^{-\frac{t}{\gamma_f t_s}(g+r\gamma_ft_s)}\Big[f_0^2\gamma_f t_s(j_++\gamma_ft_s)-g^3k_BTt_c\Big]}{gj_+j_-(g+r\gamma_ft_s)}\\
        &-\frac{f_0^2\gamma_f^2\gamma_s^2t_s^4 e^{-\frac{t}{\gamma_ft_s}(2g+r\gamma_ft_s)}}{gj_+j_-(2g+r\gamma_ft_s)}\Bigg]+\frac{f_0^2t_c^2}{j_-^2}e^{-t\big(\frac{1}{t_c}+\frac{2g}{\gamma_ft_s}\big)}\Bigg[\frac{gj_-t_c(t_s-t_c)e^{-t\big(r-\frac{2 g}{\gamma_ft_s}\big)}}{1+rt_c}+\frac{g \gamma_f \gamma_s t_c t_s^2e^{-t\big(r-\frac{g}{\gamma_ft_s}\big)}}{gt_c+(1+rt_c)\gamma_f t_s}\\
        &+\frac{2 \gamma_f^2 \gamma_s t_s^3 e^{-\frac{t}{t_c}(1-rt_c)}}{j_+}\Bigg(\frac{\gamma_s t_s}{2 g +r\gamma_ft_s}-\frac{j_- e^{\frac{gt}{\gamma_f t_s}}}{g+r\gamma_ft_s}\Bigg)\Bigg]+\frac{e^{-rt}}{r^2gj_+}\Big[-2g j_+ k_BT(g+r\gamma_s t_s)+f_0^2 t_c\Big\{2\Big[g(t_c+rt_ct_s)\\
        &+\gamma_f t_s(2+rt_s)\Big]-j_+(grt_c-2\gamma_f)+r\gamma_s^2 t_s^2\Big\}\Big]\Bigg\}+\frac{1}{g^2}e^{-rt}\Bigg\{\frac{\gamma_s t_s k_BT e^{-t\big(\frac{4 \gamma_f+3 \gamma_s}{\gamma_ft_s}\big)}}{2g(2\gamma_f+\gamma_s)}\Big[\gamma_s(4\gamma_f+3\gamma_s)e^{t\big(\frac{2\gamma_f+\gamma_s}{\gamma_ft_s}\big)}\\
        &+(8\gamma_f^2-2\gamma_s^2)e^{t\big(\frac{3\gamma_f+2\gamma_s}{\gamma_ft_s}\big)}-2g \gamma_s e^{\frac{2gt}{\gamma_ft_s}}-2g(2\gamma_f+\gamma_s)e^{\frac{3gt}{\gamma_ft_s}}-(2\gamma_f-3\gamma_s)(2\gamma_f+\gamma_s)e^{\frac{t(4\gamma_f+3\gamma_s)}{\gamma_ft_s}}\Big]+\frac{\gamma_st_s k_BT}{g}\Bigg[4\gamma_f\\
        &-4 \gamma_fe^{\frac{-gt}{\gamma_ft_s}}+\gamma_s\Big(1-e^{\frac{-2gt}{\gamma_ft_s}}\Big)-\frac{e^{-t\big(\frac{3\gamma_f+2\gamma_s}{\gamma_ft_s}\big)}}{2(2\gamma_f +\gamma_s)}\Big(\gamma_s+(2\gamma_f+\gamma_s)e^{\frac{gt}{\gamma_f t_s}}\Big)\Big(\gamma_s e^{\frac{t}{t_s}}-2 g e^{\frac{gt}{\gamma_ft_s}}+(2 \gamma_f+\gamma_s)e^{t\big[\frac{2\gamma_f+\gamma_s}{\gamma_ft_s}\big]}\Big)\Bigg]\\
        &-e^{\frac{-2gt}{\gamma_ft_s}}\big(e^{\frac{gt}{\gamma_ft_s}}-1\big)\frac{f_0^2 \gamma_s t_c t_s}{g j_+ j_-}\Big\{\gamma_f\gamma_s t_s^2+e^{\frac{gt}{\gamma_ft_s}}\Big[(4 \gamma_f +\gamma_s)\gamma_ft_s^2\Big]-2g^2t_c^2\Big\}+f_0^2t_c\Bigg[\frac{e^{\frac{-2j_+t}{\gamma_ft_c t_s}}}{j_+j_-^2}\Big(4gj_+j_-(t_c-t_s)e^{t\big(\frac{1}{t_c}\frac{2g}{\gamma_f t_s}\big)}\\
        &-g^2j_+(t_c-t_s)^2 e^{\frac{2gt}{\gamma_ft_s}}-4g\gamma_f\gamma_s t_s^2(t_c-t_s)e^{\frac{j_+t}{\gamma_f t_ct_s}}+2j_-\gamma_st_s(j_+\gamma_ft_s)e^{t\big(\frac{2}{t_c}+\frac{g}{\gamma_ft_s}\big)}-2\gamma_f\gamma_s^2t_s^3e^{\frac{2t}{t_c}}\\ 
        &+j_-e^{\frac{2j_+t}{\gamma_f t_ct_s}}\Big[t_s^2(g^2+2\gamma_f^2)-3g^2t_c^2\Big]\Big)\Bigg]\Bigg\}
    \end{split}
\end{equation}
\end{widetext}

\end{document}